\documentclass[12pt,a4paper,hidelinks]{article}
\usepackage{fancyhdr}
\usepackage{microtype}
\usepackage{authblk}
\usepackage{cite} 
\usepackage{graphicx}  
\usepackage{dcolumn}   
\usepackage{multirow}
\usepackage{bm}        
\usepackage{amssymb}   
\usepackage{amsmath}
\usepackage{booktabs}
\usepackage{epsfig}
\usepackage{tabularx}
\usepackage[subrefformat=parens,labelformat=parens,caption=false]{subfig}
\usepackage[font=small,labelfont=bf]{caption} 
\usepackage[colorlinks=false, linktocpage=true]{hyperref} 
\usepackage{rotating}
\usepackage{slashed}
\usepackage[usenames, dvipsnames]{color}
\usepackage{float}
\usepackage{framed}
\usepackage[normalem]{ulem} 
\usepackage{fancybox}

\usepackage[usenames, dvipsnames]{color}
\newcommand{\nn}{\nonumber}
\renewcommand\[{\begin{equation}} 
	\renewcommand\]{\end{equation}}
\usepackage{float}
\usepackage[normalem]{ulem} 
\usepackage[caption=false]{subfig}
\newcommand\ringring[1]{%
	{
		\mathop{\kern0pt #1}\limits^{
			\vbox to-1.85ex{
				\kern-2ex 
				\hbox to 0pt{\hss\normalfont\kern.1em \r{}\kern-.45em \r{}\hss}%
				\vss 
			}
		}
	}
}
\numberwithin{equation}{section}

\usepackage{cleveref}
\usepackage{authblk}
\begin{document}
	
	\title{\vspace{-2cm}	\textbf{Unruh-DeWitt Particle Detectors in Bouncing Cosmologies}}
	\author[1]{Aindri\'u Conroy \thanks{aindriu.conroy@matfyz.cuni.cz}}
	\affil[1]{\textit{Institute of Theoretical Physics, Faculty of Mathematics and Physics,
			Charles University, V Hole\v{s}ovi\v{c}k\'ach 2, Prague, 18000, Czech Republic.}\vspace{.5cm}}
	\author[2]{Peter Taylor \thanks{peter.taylor@dcu.ie}}
	\affil[2]{\textit{Centre for Astrophysics and Relativity,
			School of Mathematical Sciences,
			Dublin City University,
			Glasnevin,
			Dublin 9, Rep. of Ireland.}
	}
	\date{\today}
	\maketitle
 \flushbottom
	\begin{abstract}
		We study semi-classical particle production in non-singular bouncing cosmologies by employing the Unruh-DeWitt model of a particle detector propagating in this class of spacetimes. The scale factor for the bouncing cosmology is derived analytically and is inspired by the modified Friedmann equation employed in the \emph{loop quantum cosmology} literature. We examine how the detector response varies with the free parameters in this model such as the equation of state during the contraction phase and the critical energy density during the bounce phase. We also investigate whether such a signature in the particle detector survives at late times.
	\end{abstract}
	\newpage
	\tableofcontents
	\section{Introduction}
	
	Despite its enormous success, Einstein's theory of \emph{general relativity} points to its own shortcomings at the smallest length scales. That is to say that in the  cosmological context, provided certain energy conditions are satisfied, solutions of the Einstein equations necessarily result in a singularity at a finite time in the past. This is the so-called \emph{big bang} singularity. From a purely classical viewpoint,  if we  wish to circumvent this singularity, there are two options available to us, either we modify the Einstein equations or we violate one or more of the energy conditions, the latter of which constitute important assumptions in the Hawking-Penrose singularity theorems \cite{Hawking:1973uf}.
	
	Leaving aside the steady-state models, such as those proposed by Bondi and Gold in Refs.~\cite{Bondi:1948qk,Bondi:1954qj}, which have been found to be inconsistent with the observed properties of the \emph{cosmic microwave background} (CMB), non-singular models can largely be categorised according to the \emph{bounce mechanism} they invoke to resolve the initial singularity. This mechanism can be sourced from some modified theory of gravity, such as through the introduction of higher-derivative corrections to the gravitational action which, under certain conditions, can resolve the singularity, see for instance the defocusing conditions detailed in Ref.~\cite{Conroy6}; or from the introduction of (potentially exotic) matter as in the \emph{matter bounce scenario} discussed by Brandenburger et al. in Refs.~\cite{Branden2002,Brandenberger:2012zb}, or indeed energy condition violating models such as that proposed by Steinhardt and Ijjas in Refs.~\cite{IJJAS2017289,Ijjas:2018qbo}. 
 
 If the characteristic length scale associated with the bounce is sufficiently small -- by which we mean, length scales that approach the Planck length -- then it is not possible to ignore the quantum effects of spacetime itself. One proposal for describing the quantum nature of spacetime is \emph{loop quantum gravity} (LQG), and it is claimed that working in this framework does indeed avoid a \emph{big bang} singularity, producing instead a quantum bounce \cite{Agullo:2016tjh,Rovelli:2014ssa}. More importantly for our present consideration is that there exists a well-defined semi-classical limit to quantum theory in LQG \cite{Taveras:2008ke} and moreover, this limit can be expressed in a surprisingly accurate effective Friedmann equation that includes leading-order quantum corrections. Technically speaking, the semi-classical limit and the associated effective Friedmann equation are only valid at late times, but it was shown in Ref.~\cite{Rovelli:2013zaa} to be a robust approximation well outside of its formal domain of validity, including early times where the equations can predict a bounce rather than a singularity. Notwithstanding these technical subtleties, we simply take the effective Friedmann equation derived from the semi-classical limit of LQG as our governing equation for the dynamics of an effective classical spacetime at early times. Indeed it is essential to the self-consistency of our own set-up that we are not in a regime where the quantum effects of spacetime become important. This assumption is implicit in the adoption of the Unruh-DeWitt particle detector model which assumes the framework of \emph{quantum field theory in curved spacetimes} (QFTCS).
	
	In this paper, we study the response of an Unruh-DeWitt particle detector propagating in a class of bouncing cosmologies, examining how the detector response varies with the free parameters in the model such as the equation of state during the contraction phase and the magnitude of the correction in the effective Friedmann equation. We also compare the detector response with the response seen in a \emph{big bang} cosmology. One might expect that, since the detector is localised in spacetime, it would be quite agnostic as to whether the universe began with a bounce or a bang, but since the detector's response depends on its entire history and on the quantum state of a quantized scalar field, it encodes global information about the spacetime. As such, we use this structure to ask what can a study of semi-classical particle production tell us about the early universe. We have three principal avenues of investigation: (i) Does the transition rate of a detector exhibit any robust general features as it traverses the bounce? (ii) Is the equation of state parameter for the contraction phase imprinted on the transition rate at late times? (iii) Are \emph{big bang} cosmologies distinguishable from bouncing cosmologies in the transition rate at late times?
	
	The paper is organised as follows: in Section~\ref{sec:model}, we develop an analytic cosmological model by solving a modified Friedmann equation for the bounce phase and the standard Friedmann equation for all other eras, and then pasting the solutions together with appropriate boundary conditions. In Section~\ref{sec:detector}, we review some general details of the Unruh-DeWitt particle detector model as well as discussing its calculation in the current context. In Section~\ref{sec:results}, we present a detailed survey of our results with regard to answering the questions set out above. 
~
 \section{Analytic model}
	\label{sec:model}
	In this section, we present details of the cosmological model that we adopt for our investigation. We assume a flat FLRW spacetime with line element (in spherical polar coordinates)
	\begin{equation}
		\label{eq:FLRW}
		ds^{2}=-dt^{2}+a^{2}(t)\left(dr^2+r^{2}d\theta^{2}+r^{2}\sin^{2}\theta\,d\phi^{2}\right),
	\end{equation}
	where we look for analytical approximations for the scale factor $a(t)$ in different cosmological eras. In particular, we assume a universe with four distinct eras: (i) a contraction phase, (ii) a radiation-dominated bounce phase, (iii) a matter domination phase, and (iv) a dark-energy domination phase which takes over at late times.

	During the contraction phase, the equation of state parameter $w=p/\rho$ is a free parameter in the model and so we investigate how different choices manifest in the detector's response. Indeed, we devote Section \ref{sec:EOS} to this particular study.
 
		The other free parameter in the model comes from the bounce phase which emerges from a modification of the Friedmann equation in the radiation-dominated era. This model is inspired by the bouncing models of \emph{loop quantum gravity} as described in Ref.~\cite{Rovelli:2013zaa,Taveras:2008ke}, though again we emphasize that we are taking a phenomenological approach untethered to any particular bounce mechanism. Indeed, as mentoned in the introduction, it is important for the self-consistency of our Unruh-DeWitt detector model that we treat the background spacetime classically. The freedom in the bounce phase comes from the critical energy density $\rho_{cr}$ which parametrizes the deviation from the \emph{big bang} model. When $\rho_{cr}\to\infty$, we retrieve the standard \emph{big bang} cosmology. 
	
	Beyond the bounce phase, our strategy is the simplest one, our analytical approximation in each era is the solution to the standard, classical Friedmann equation with only the dominant energy density contributing to that era. We note, however, that the detector's response at any instant will depend on the scale factor from any prior eras that the detector has passed through since the detector's response depends on the detector's entire history, as discussed further in Section~\ref{sec:detector}.

	\subsection{Contraction phase}
	In the contraction phase, we assume the equation of state  $p(t)=w\,\rho(t)$ where $w$ is a constant, and solve the  continuity equation to give
	\begin{equation}
		\label{continuityeq}
		\rho(t)=\rho_{r}\left(a(t)/a_{r}\right)^{-3(w+1)}.
	\end{equation}
	Here, we have introduced $\rho_{r}\equiv \rho(t_{r})$ and $a_{r}\equiv a(t_{r})$ where $t_{r}<0$ is the time at which the contraction phase transitions to the radiation-dominated bounce phase. To obtain the scale factor as a function of time, we solve the Friedmann equation
	\begin{equation}
		\label{contFried}
		H^{2}(t)=\frac{\kappa\rho(t)}{3},
	\end{equation}
	where we denote $\kappa=8\pi G$ for convenience. The solution is
	\begin{equation}
		a(t)=a_{r}\left(1-\frac{\sqrt{3\kappa\rho_{r}}}{2}(w+1)(t-t_{r})\right){}^{\frac{2}{3(w+1)}}.
	\end{equation}
	
	\subsection{Radiation-dominated bounce phase}
	Turning now to the scale factor in the  era which contains the bounce. We assume the energy density is dominated by radiation with equation of state $p(t)=\tfrac{1}{3}\rho(t)$, whence the continuity equation implies $\rho(t)\propto a^{-4}(t)$. The standard Friedmann equation with this energy density produces a singularity at a finite time in the past, the so-called \emph{big bang} singularity. Hence, we are compelled to modify this equation if we are to avoid this pathology. We consider the following modified Friedmann equation
	\begin{equation}
		\label{FriedQu}
		H^{2}(t)=\frac{\kappa}{3}\rho(t)\left(1-\frac{\rho(t)}{\rho_{cr}}\right),
	\end{equation}
	where $\rho_{cr}$ is the constant critical energy density which  measures the deviation from the singular solution. When $\rho(t)\ll\rho_{cr}$,  the dynamics of the spacetime is well approximated by the standard Friedmann equation. This modification is derived from the semi-classical limit of \emph{loop quantum gravity} applied to quantising cosmological spacetimes, see for example Ref.~\cite{Taveras:2008ke}, and moreover it has been argued in Ref.~\cite{Rovelli:2013zaa} that this semi-classical limit, while formally only valid at late times, is surprisingly accurate outside of its formal domain of validity including being able to predict a bounce rather than a singularity. Regardless of its origin in loop quantum cosmology, here we adopt a phenomenological approach by taking this equation as our governing equation for the scale factor during this phase.
	
	Solving Eq.~(\ref{FriedQu}) is straightforward and results in the scale factor
	\begin{equation}
		\label{ar0}
		a(t)=a_{0}\left[1+\frac{4}{3}\sqrt{3\kappa\rho_0(1-\rho_{0}/\rho_{cr})}t+\frac{4\kappa\rho_{0}}{3}t^{2}\right]^{1/4},
	\end{equation}
	where $a_{0}\equiv a(0)$ and $\rho_{0}\equiv \rho(0)$. For a bounce to occur at $t=0$, then this must be a local minimum of the scale factor which requires
	\begin{equation}
		\label{bouncecond}	H(t)\bigr|_{t=0}=0\quad\mbox{and}\quad \dot{H}(t)\bigr|_{t=0}>0.
	\end{equation}
	This, in turn, implies that
	$\rho_{0}=\rho_{cr}$ yielding the following solution during the bounce phase 
	\[
	\label{Rada}
	a(t)=a_{0}\left[1+\frac{4\kappa\rho_{cr}}{3}t^{2}\right]^{1/4}.
	\]
	Since $\rho(t)\propto a^{-4}(t)$, $t=0$ is a maximum of the energy density and hence $\rho_{cr}=\rho_{0}$ can be interpreted as the maximum energy density.

	\subsection{Matter domination phase}
	In the matter-dominated era, which follows the bounce phase at time $t_m$, we take $w=0$ to find $\rho(t)\propto a^{-3}(t)$ . Solving the standard Friedmann equation then yields
	\begin{equation}	a(t)=a_{m}\left(1+\frac{\sqrt{3\kappa\rho_{m}}}{2}(t-t_{m})\right)^{2/3},
	\end{equation}
	where $a_{m}\equiv a(t_{m})$ and $\rho_{m}=\rho(t_{m})$.
	
	\subsection{Dark Energy domination phase}
	At late times, the universe evolves to be dominated by dark energy,  which is the special case of $p(t)=-\rho(t)$.  In this case, the only term on the righthand side of the Friedmann equation is the cosmological constant term
	\begin{equation}
		\frac{\dot{a}}{a}=\sqrt{\frac{\Lambda}{3}}\quad\implies\quad a(t)\propto  e^{\sqrt{\frac{\Lambda}{3}}t},
	\end{equation}
	which is, of course, the scale factor for de Sitter space.
	As this era contains the present day, we introduce
	\begin{equation}
		\label{LambdaBar}
		\bar\Omega_\Lambda=\frac{\Lambda}{3\bar H^{2}},
	\end{equation}
	where `barred' quantities signify that they have been evaluated today so that $\bar\Omega_\Lambda$ is the present value of the density parameter for dark energy. This allows us to write
	\begin{equation}
		\label{atLambda}
		a(t)=\bar{a}_{\Lambda}e^{\bar {H}\sqrt{\bar{\Omega}_{\Lambda}}(t-t_{\Lambda})}.
	\end{equation}
	
	\subsection{Smooth boundary conditions}
	Having attained the dominant form of the scale factor over the entire
	evolution of the universe, we must now set about tuning the free integration
	constants in the model to ensure that the scale factor is smooth. In the first instance, we represent the scale factor by the piecewise function
	\begin{equation}
		a(t)=\begin{cases}
			a_{r}\left(1-\frac{\sqrt{3\kappa\rho_{r}}}{2}(w+1)(t-t_{r})\right){}^{\frac{2}{3(w+1)}} & t<t_{r}\\
			a_{0}\left(1+\frac{4\kappa\rho_{cr}}{3}t^{2}\right)^{1/4} & t_{r}\leq t<t_{m}\\
			a_{m}\left(1+\frac{\sqrt{3\kappa\rho_{m}}}{2}(t-t_{m})\right){}^{2/3} & t_{m}\leq t<t_{\Lambda}\\
			a_{\Lambda}\;e^{\bar{H}\sqrt{\bar{\Omega}_{\Lambda}}(t-t_{\Lambda})} & t\geq t_{\Lambda}.
		\end{cases}
	\end{equation}
	If we label the scale factor during the bounce phase to be
 \begin{align}
     a_{b}(t)=a_{0} (1+\tfrac{4}{3}\kappa\,\rho_{cr} t^{2})^{1/4},
 \end{align}
 then continuity in the scale factor implies
	\begin{align}
		\label{eq:continuity}
		a_{r}&=a_{b}(t_{r}),\quad a_{m}=a_{b}(t_{m}), \quad a_{\Lambda}=a_{m}\left(1+\frac{\sqrt{3\kappa\rho_{m}}}{2}(t_{\Lambda}-t_{m})\right)^{2/3},
	\end{align}
	while continuity in the derivative implies
	\begin{align}
		\rho_{r}&=\frac{4}{3}\kappa\,\rho_{cr}^{2}t_{r}^{2}\left(\frac{a_{0}}{a_{r}}\right)^{8},\qquad \rho_{m}=\frac{4}{3}\kappa\,\rho_{cr}^{2}t_{m}^{2}\left(\frac{a_{0}}{a_{m}}\right)^{8},\nonumber\\	\bar{H}\sqrt{\bar{\Omega}_{\Lambda}}&=\frac{2\kappa\,\rho_{cr}t_{m}a_{0}^{4}}{3(a_{m}^{4}+\kappa\,\rho_{cr}t_{m}a_{0}^{4}(t_{\Lambda}-t_{m}))}.
	\end{align}
	We note here that, by using (\ref{eq:continuity}), we can express $a_{\Lambda}$ in terms $a_{0}$ and $\rho_{cr}$ like so
	\begin{align}
a_{\Lambda}=a_{m}\left(1+\kappa\,\rho_{cr}t_{m}(a_{0}/a_{m})^{4}(t_{\Lambda}-t_{m})\right)^{2/3}.
	\end{align}

 The critical energy density is a free parameter in the model and we would like to explore the consequences of different choices of $\rho_{cr}$ in the response of the particle detector. It is preferable to employ dimensionless free parameters as different choices will not depend on the particular units involved. As $\kappa\rho_{cr}$ has dimension $[T]^{-2}$ where $T$ is time, we will define the dimensionless critical energy density $\tilde{\rho}_{cr}$ in terms of the duration of the bounce phase, i.e. $\tilde{\rho}_{cr}=\kappa\rho_{cr}\,(t_{m}-t_{r})^{2}$. If we also define $\tilde{t}=t/(t_{m}-t_{r})$, then the scale factor in terms of $\tilde{t}$ can be expressed as
 \begin{align}
 \label{eq:scalefactor}
     a(\tilde{t})=\begin{cases}
         a_{r}\left(1-\left(\frac{a_{0}}{a_{r}}\right)^{4}\tilde{\rho}_{cr}|\tilde{t}_{r}|(1+w)(\tilde{t}-\tilde{t}_{r})\right)^{\tfrac{2}{3(1+w)}}\qquad &\tilde{t}<\tilde{t}_{r}\\
         a_{0}(1+\tfrac{4}{3}\tilde{\rho}_{cr}\tilde{t}^{2})^{1/4}  & \tilde{t}_{r}\le \tilde{t}<\tilde{t}_{m}\\
         a_{m}\left(1+\left(\frac{a_{0}}{a_{m}}\right)^{4}\tilde{\rho}_{cr}\tilde{t}_{m}(\tilde{t}-\tilde{t}_{m})\right)^{2/3} & \tilde{t}_{m}\le t<\tilde{t}_{\Lambda}\\
         a_{\Lambda}\exp\left(\frac{2\tilde{\rho}_{cr}\tilde{t}_{m}a_{0}^{4}(\tilde{t}-\tilde{t}_{\Lambda})}{3 a_{m}^{5/2}a_{\Lambda}^{3/2}}\right) & \tilde{t}\ge \tilde{t}_{\Lambda},
     \end{cases}
 \end{align}
 where it is helpful to express the scale factors at the transitions in terms of $\tilde{\rho}_{cr}$,
 \begin{align}
     a_{r}&=a_{0}(1+\tfrac{4}{3}\tilde{\rho}_{cr}\tilde{t}_{r}^{2})^{1/4},\qquad  a_{m}=a_{0}(1+\tfrac{4}{3}\tilde{\rho}_{cr}\tilde{t}_{m}^{2})^{1/4},\nonumber\\
     a_{\Lambda}&=a_{m}^{-5/3}(a_{m}^{4}+a_{0}^{4}\tilde{\rho}_{cr}\tilde{t}_{m}(\tilde{t}_{\Lambda}-\tilde{t}_{m}))^{2/3}.
 \end{align}
With this rescaling, it is clear that the only interesting free parameter in the model is the dimensionless critical energy density. The total bounce phase is equal to unity in these units. Throughout the remainder of the paper, we take $\tilde t_r=-1/2$ and $\tilde t_m=1/2$. The normalization of the scale factor at the bounce $a_{0}$ is irrelevant, we take it to be unity henceforth.
	
	\section{The Unruh-DeWitt particle detector}
	\label{sec:detector}
	As a mathematical model for our particle detector, we consider the Unruh-DeWitt model introduced in Ref.~\cite{Unruh:1976} which centres on a two-level idealized atom interacting with a massless quantum scalar field. Absorption of field quanta by the atom promotes the atom from ground state to excited state and we interpret this atomic excitation as a detector registering a particle. Conversely, the detector can de-excite by emitting quanta.
	
	Before the detector and the quantum field interact, we suppose that the field $\hat\varphi(x)$ is in some initial Hadamard state $|\Phi_{\textrm{i}}\rangle$ on a given background, while the detector is in an initial energy eigenstate $|E_{\textrm{i}}\rangle$. When interaction takes place, the field $\hat\varphi(x)$ transitions from its initial state $|\Phi_{\textrm{i}}\rangle$ to a final state $|\Phi_{\textrm{f}} \rangle$, while the detector undergoes a transition from initial eigenstate $|E_{\textrm{i}}\rangle$ to its other eigenstate $|E_{\textrm{f}}\rangle$. The probability of this transition occurring will not depend on the individual eigenvalues of these energy states but only on the difference $E=E_{\textrm{f}}-E_{\textrm{i}}$, which we call the \emph{energy gap}. When $E>0$, we are considering the probability that the detector has absorbed a field quanta and is in an excited state, while for $E<0$ we are considering the probability the detector has de-excited by emitting a field quanta. If one assumes a weak coupling between the detector and the quantum field, then one can treat the interaction perturbatively. To first order in this perturbative expansion, and tracing over the field degrees of freedom -- since we are only interested in measuring the detector's state -- the probability of measuring the detector in an excited state $|E_{\textrm{f}}\rangle$ is proportional (the constant of proportionality being irrelevant\footnote{Explicitly the probability function is $P(E)=\alpha^2 |\langle E_f|\hat\mu(0)|E_i\rangle|^2 {\cal F}(E)$, where $\alpha$ is a coupling constant (assumed to be small) and $\hat \mu(\tau)$ is the detector's monopole moment operator. The constant of proportionality $\alpha^2 |\langle E_f|\hat\mu(0)|E_i\rangle|^2$ depends only on the detector's internal structure and is independent of the detector's trajectory and the quantum state of the field.}) to the response function
	\begin{align}
		\label{eq:response}
		\mathcal{F}(E)=2&\,\lim_{\epsilon\to 0^{+}} \Re \int_{-\infty}^{\infty}du\,\chi(u)\int_{0}^{\infty}ds\,\chi(u-s)e^{-i \,E\,s}W_{\epsilon}(u, u-s),
	\end{align}
	where $\chi(s)\equiv \chi(x(s))$ is the switching function which dictates how the interaction is switched on and the bi-distribution $W_{\epsilon}(u,u-s)\equiv W_{\epsilon}(x(u),x(u-s))$ appearing in this integral is a one-parameter family of Wightman Green functions for the quantum scalar field evaluated at the spacetime points $x=x(u)$ and $x'=x(u-s)$. There is an implicit `$i\,\epsilon$' prescription in this expression which is required to render the Wightman Green function a well-defined distribution on the lightcone, see, for example, Ref.~\cite{BirrellDavies} for further details. It is also assumed that the switching function $\chi$ is smooth and of compact support so that the integral above is well-defined.  
	
	For practical computations, it is preferable to have an explicit regularisation for the Wightman Green function. This is possible provided the field is in a quantum state that satisfies the Hadamard condition \cite{WaldBook}, where the response function is given by
	\begin{align}
		\label{responsereg}
		\mathcal{F}&(E)  =
		-\frac{E}{4\pi}\int_{-\infty}^{\infty}\chi^{2}(u)du+\frac{1}{2\pi^{2}}\int_{0}^{\infty}ds\frac{1}{s^{2}}\int_{-\infty}^{\infty}du\,\chi(u)\left[\chi(u)-\chi(u-s)\right]\nonumber\\
		&+2\int_{-\infty}^{\infty}du\,\chi(u)\int_{0}^{\infty}ds\,\chi(u-s)
		\left(\cos (E s)\;W(u,u-s)+\frac{1}{4\pi^{2}s^{2}}\right),
	\end{align}
 as in Ref.~\cite{LoukoSatz2008}, where $W(u,u-s)=\lim_{\epsilon\to 0}W_{\epsilon}(u,u-s)$. Now, $W(u,u-s)$ is a well-defined distribution everywhere except at the vertex of the lightcone (when $x(u)=x(u-s)$ or equivalently when $s=0$) but this pathology is now explicitly regularised by the counterterm $1/(4 \pi^{2}s^{2})$. 
	
	In our investigation here, we wish to consider the case of sharp-switching where the detector is switched on and off instantaneously. While this violates the assumption that the switching function is smooth, it is possible to consider sharp-switching as the limit of an increasingly steep smooth switching function. The resulting response function contains a logarithmic term which diverges at this limit. Nevertheless, the \emph{rate} of this response function is regular even at the sharp-switching limit and is given by \cite{LoukoSatz2008}
	\begin{align}
		\label{eq:transitionratesharp}
		\dot{ \mathcal{F}}_{\tau}(E)=
		2\int_{0}^{\Delta\tau}ds\,\left(\cos(E\, s)\;W(\tau,\tau-s)+\frac{1}{4\pi^{2}s^{2}}\right)-\frac{E}{4\pi}+\frac{1}{2\pi^{2}\Delta\tau},
	\end{align}
where the notation $\dot{ \mathcal{F}}_{\tau}$ indicates a derivative of $\mathcal{F}$ with respect to proper time $\tau$. Further details of sharp-switching limit can be found in Appendix~\ref{sec:Sharp} and Refs.~\cite{Satz2007,Conroy14}.
 
	This is the quantity of primary interest for the remainder of this article. Again, we note that while the above is technically proportional to the transition rate, for simplicity we will refer to it as the transition rate for the remainder of the article. Hence Eq.~(\ref{eq:transitionratesharp}) represents (up to a constant of proportionality) the number of excitations (or de-excitations) per unit time in an ensemble of identical detectors \cite{LoukoSatz2008}.
 The integral above is then tantamount to integrating the (regularised) Wightman Green function for the field over the worldline of the detector with a weighting that depends on the energy gap of the detector's states. Now it is important to qualify that, while the transition probability itself must be positive, the transition rate can take on negative values. If we measure the system at some time $\tau$, then the detector will be in some superposition of its ground and excited state, with some coefficients representing the amplitudes of each state. As a function of the measurement time, these amplitudes need not be monotonic in $\tau$ and in particular, they can destructively interfere in such a way that the probability of excitation (or de-excitation) decreases as a function of time. This corresponds to a negative transition rate, as explained in more detail in, for example, Ref.~\cite{Foo_2020}.
	
	Turning now to our particular cosmological context. The FLRW spacetime (\ref{eq:FLRW}) is conformal to Minkowski spacetime. If we further take our quantum scalar field to be conformally invariant by assuming the field is conformally coupled to the background geometry and that the field is in the conformal vacuum state \cite{BirrellDavies}, then the Wightman two-point function is conformally related to the Minkowski two-point function in conformal coordinates. Namely, we define conformal time
	\begin{align}
 \label{eq:conftdef}
		\eta=\int^{t}\frac{1}{a(t')}dt'
	\end{align}
	whence the metric is explicitly conformally flat $ds^{2}=a^{2}(\eta)\left[-d\eta^{2}+d\textbf{x}^{2}\right]$ and the two-point function is
	\begin{align}
		W(x,x')=\frac{1}{4\pi^{2}a(\eta)a(\eta')}\frac{1}{(-\Delta\eta^{2}+\Delta\textbf{x}^{2})}.
	\end{align}
 Explicitly, if we define a dimensionless conformal time via $\tilde{\eta}=\eta/(t_{m}-t_{r})$, we can express conformal time as the piecewise function
 \begin{align}
 \label{eq:etaDef}
\tilde{\eta}=\begin{cases}
\displaystyle{\frac{3a_{r}^{3}}{a_{0}^{4}(3w+1)\tilde{\rho}_{cr}|\tilde{t}_{r}|}\left[1-\Big(1-\frac{a_{0}^{4}}{a_{r}^{4}}(1+w)\tilde{\rho}_{cr}|\tilde{t}_{r}|(\tilde{t}-\tilde{t}_{r})\Big)^{\tfrac{1+3w}{3(1+w)}}\right]+F(\tilde{t}_{r})},  \\
\qquad\qquad\qquad\qquad\qquad\qquad\qquad\qquad\qquad\qquad\qquad\qquad\qquad\quad \tilde{t}<\tilde{t}_{r}\\ \\
\displaystyle{F(\tilde{t})}, \qquad\qquad\qquad\qquad\qquad \qquad\qquad\qquad\qquad\qquad\quad\quad\,\, \tilde{t}_{r}\le\tilde{t}<\tilde{t}_{m}\\ \\
\displaystyle{\frac{3a_{m}^{3}}{a_{0}^{4}\tilde{\rho}_{cr}\tilde{t}_{m}}\Big[-1+\Big(1+\frac{a_{0}^{4}}{a_{m}^{4}}\tilde{\rho}_{cr}\tilde{t}_{m}(\tilde{t}-\tilde{t}_{m})\Big)^{\tfrac{1}{3}}\Big]+F(\tilde{t}_{m})}, \quad\quad\,\,\tilde{t}_{m}\le\tilde{t}<\tilde{t}_{\Lambda}\\ \\
\displaystyle{\frac{3a_{m}^{5/2}a_{\Lambda}^{1/2}}{a_{0}^{4}\tilde{\rho}_{cr}\tilde{t}_{m}}\left[\frac{3}{2}-\frac{a_{m}^{1/2}}{a_{\Lambda}^{1/2}}-\frac{1}{2}\exp\left(\frac{-2\tilde{\rho}_{cr}\tilde{t}_{m}a_{0}^{4}(\tilde{t}-\tilde{t}_{\Lambda})}{3 a_{m}^{5/2}a_{\Lambda}^{3/2}}\right)\right]+F(\tilde{t}_{m})},\\ \qquad\qquad\qquad\qquad\qquad\qquad\qquad\qquad\qquad\qquad\qquad \qquad\qquad\,\,\,\,\tilde{t}\ge\tilde{t}_{\Lambda}
\end{cases}
 \end{align}
 where
 \begin{align}
     F(\tilde{t})\equiv \frac{\tilde{t}}{a_{0}}\,{}_{2}F_{1}\left(\tfrac{1}{4},\tfrac{1}{2},\tfrac{3}{2},-\tfrac{4}{3}\tilde{\rho}_{cr}\tilde{t}^{2}\right),
 \end{align}
 with ${}_{2}F_{1}(a,b,c,z)$ the standard Hypergeometric function. The constants of integration in (\ref{eq:conftdef}) are chosen so that the conformal time is a smooth function of $\tilde{t}$ and it is zero when $\tilde{t}=0$.
 
	For simplicity, we will focus only on the case of a comoving detector $\{t=\tau,\textbf{x}=const.\}$ for which the transition rate is
	\begin{align}
		\label{eq:transitionratecomoving}
		\dot{ \mathcal{F}}_{\tau}(E)=
		\frac{1}{2\pi^{2}}\int_{0}^{\Delta\tau}ds\,\left(-\frac{\cos(E\, s)}{a(\tau)a(\tau-s)\left[\eta(\tau)-\eta(\tau-s)\right]^{2}}+\frac{1}{s^{2}}\right)\nonumber\\
		-\frac{E}{4\pi}+\frac{1}{2\pi^{2}\Delta\tau}.
	\end{align}
	In principle, computing (\ref{eq:transitionratecomoving}) ought to be straightforward since we know the scale factor in closed form. However, there is an additional obstacle associated with the natural units in which we are working. In units where $G=c=\hbar=1$, the interval in the integral above will typically be very large while at the same time, reasonable values for the energy gap in the integral are very small. This presents a numerical difficulty with the evaluation of the integral above. We can circumvent this issue by rescaling the integration variable by $s\to\Delta\tau\,s$ and using the dimensionless (proper) time $\tilde{\tau}=\tau/(t_{m}-t_{r})$ that we adopted above, to obtain
 \begin{align}
 \label{eq:FDotTilde}
     \dot{\tilde{\mathcal{F}}}_{\tau}(E)=\frac{1}{2\pi^{2}\Delta\tilde{\tau}}\int_{0}^{1}ds\Bigg[-\frac{\cos(\tilde{E}\,\Delta\tilde{\tau}\,s)}{a(\tilde{\tau})a(\tilde{\tau}-\Delta\tilde{\tau}\,s)}&\left(\frac{\Delta\tilde{\tau}}{\tilde{\eta}(\tilde{\tau})-\tilde{\eta}(\tilde{\tau}-\Delta\tilde{\tau}\,s)}\right)^{2}+\frac{1}{s^{2}}\Bigg]\nonumber\\
    & -\frac{\tilde{E}}{4\pi}+\frac{1}{2\pi^{2}\Delta\tilde{\tau}},
 \end{align}
where we have defined the dimensionless transition rate $\dot{\tilde{\mathcal{F}}}_{\tau}(E)\equiv (t_{m}-t_{r})\dot{\mathcal{F}}_{\tau}(E)$ and the dimensionless energy gap $\tilde{E}\equiv (t_{m}-t_{r})E$. Everything in (\ref{eq:FDotTilde}) is now expressed in terms of dimensionless quantities and will be agnostic to the units we choose.

Eq.~(\ref{eq:FDotTilde}) is now readily computable. We numerically computed the integral above in Mathematica using the explicit expressions for the scale factor and conformal time in (\ref{eq:scalefactor}) and (\ref{eq:etaDef}), respectively. The numerical integrals were computed with a working precision of 64 decimal places and an absolute accuracy of 12 decimal places. While the integral is regular at $s=0$, the cancellation of the singularities at $s=0$ between the Wightman function and the $1/s^{2}$ terms is not resolvable numerically. Instead we cut the integral off very close to zero, at $10^{-16}$, well below our overall accuracy goal. The only other numerical consideration that we highlight is that plotting $\dot{\tilde{\mathcal{F}}}_{\tau}(E)$ as a function of proper time involves computing (\ref{eq:FDotTilde}) on a grid of $\tau$ values and interpolating between them. When the energy gap is large, the frequency of oscillations is large and the $\tau$ grid must be very finely meshed in order to resolve the oscillations.

We finish this section with a brief explanation of why we obtain non-trivial results in the detector response given that Parker proved in his seminal work in the 1960s that there is no real particle production for a conformally invariant quantized scalar field ~\cite{parker1967creation,Parker1968}. There are two essential distinctions between the framework adopted here and the one used in standard cosmological particle production. The first is the notion of a particle itself. In the standard approach to cosmological particle production (see Ref.~\cite{Parker:2012at} for a recent review), one (usually) models the universe with two asymptotically flat regions in the distant past and future where the ordinary QFT notion of a particle exists. One can compare the particle content of the vacua in these asymptotic regions via Bogolubov transformations to ascertain whether particles have been created during the intermediate dynamic evolution of the universe. Contrariwise, the Unruh-DeWitt model which we adopt herein recognizes that there is no unambiguous notion of a particle in curved spacetime and we simply define a particle \textit{operationally} through the maxim of ``a particle is what a particle detector detects''. This operational meaning is attributed to Bill Unruh in Ref.~\cite{Fewster:2015dqb}. The second important distinction is that Parker's proof that there is no particle production for the conformally invariant field, assumes that the quantized field is a free field. However, in the Unruh-DeWitt framework, the field is interacting (albeit weakly) with the detector. The Unruh effect \cite{Unruh:1976} is a good example of non-trivial particle detection arising from this interaction between the field and the particle detector in an otherwise trivial spacetime.

	\section{Results}
	\label{sec:results}
 \subsection{Quantum effects near the bounce}
 	\label{sec:Early times}
  \begin{figure}[htb!]
			\centering
			\resizebox{\linewidth}{!}{%
				\subfloat[\label{fig:SPA1} Bounce phase throat]{
					\includegraphics[height=4.7cm]{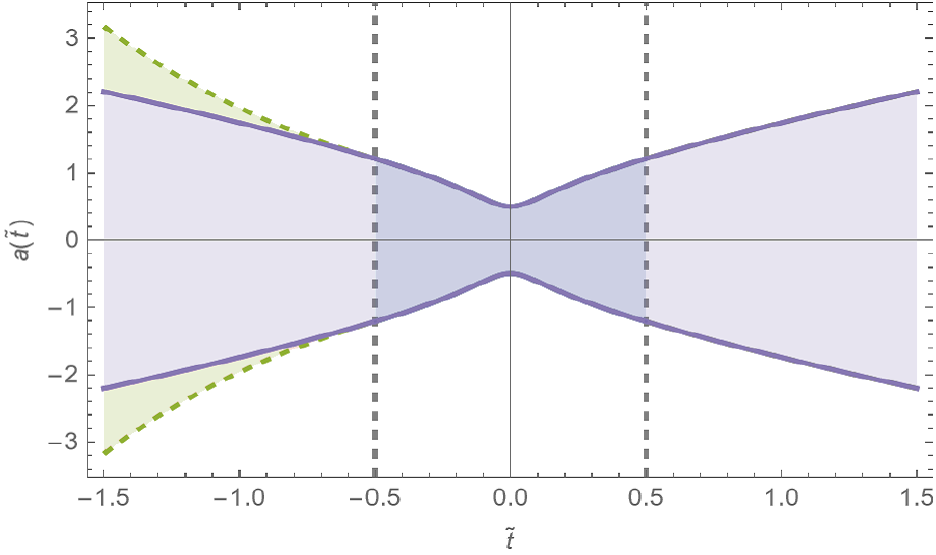}
				}
				\subfloat[\label{fig:SPA2}Scale factor at early times]{
					\includegraphics[height=4.7cm]{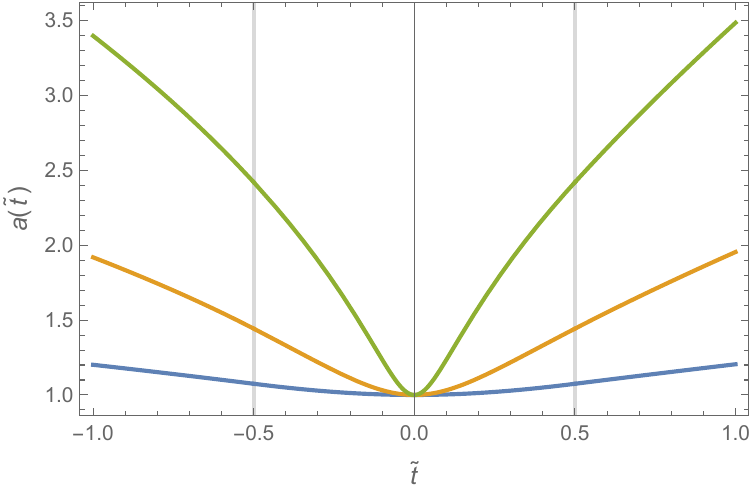}	
			}}
			\\	\resizebox{\linewidth}{!}{
				\subfloat[\label{fig:SPA3} Hubble parameter at early times]{
					\includegraphics[height=4.7cm]{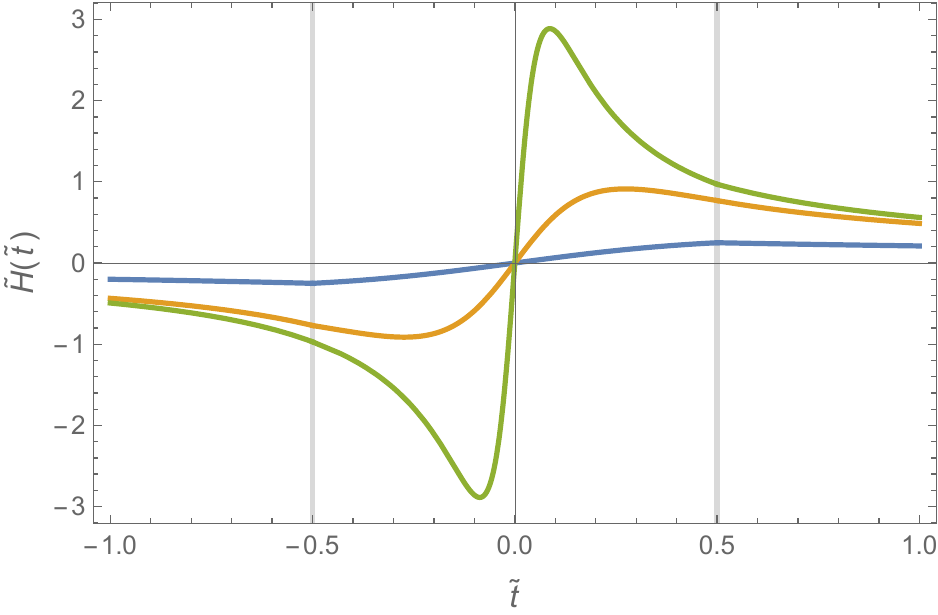}
				}
				\subfloat[\label{fig:SPA4} Hubble radius $1/|\tilde H(\tilde t)|$]{
					\includegraphics[height=4.7cm]{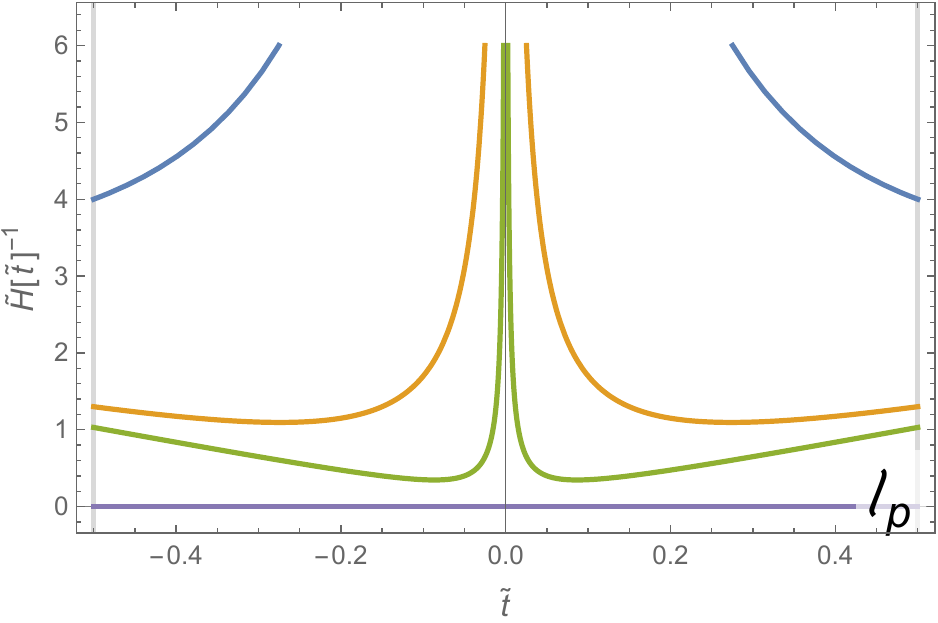}
				}}
				\caption{In Fig.~\protect\subref*{fig:SPA1}, we plot the scale factor through the bounce phase (the blue shaded region). Different contraction equations of state produce different shape throats during the contraction phase. The gray dashed lines at $\tilde{t}=\pm 1/2$ correspond to the transition between the contraction and bounce phase and between the bounce phase and matter-dominated era. In each of the remaining figures we plot the scale factor, Hubble parameter and Hubble radius for bouncing models with different values of the critical energy density. The blue curve corresponds to $\tilde \rho_{cr}=1$, the orange curve corresponds to $\tilde \rho_{cr}=10$ and the green curve is $\tilde \rho_{cr}=100$. Figure~\protect\subref*{fig:SPA2} depicts the scale factor around the bounce;  Fig.~\protect\subref*{fig:SPA3} illustrates the sign change the Hubble parameter undergoes across the bounce; while, in  Fig.~\protect\subref*{fig:SPA4}, we compare the Hubble radius for these models with that of the Planck length.}
    			\label{fig:SpacetimeA}
			\end{figure}
 In this subsection, we present results for how an Unruh-DeWitt detector responds as it traverses the bounce. Before discussing the numerical results however, we will first spend some time considering what behaviour we would intuitively expect to observe. To aid with this, we turn to Fig.~\ref{fig:SpacetimeA} where we plot the scale factor, the (dimensionless) Hubble parameter,  and the (dimensionless) Hubble radius for different values of the (dimensionless) critical energy density $\tilde{\rho}_{cr}$. 
 
 Let us first consider Fig.~\subref*{fig:SPA4} where we plot the Hubble radius around the bounce. At a fixed time $\tilde{t}$, we can think of $|\tilde{H}(\tilde{t})|^{-1}=(t_{m}-t_{r})^{-1}|H(\tilde{t})|^{-1}$ as a measure of the importance of quantum gravity effects, the smaller this quantity the more important quantum gravity effects become. Equivalently, we can largely ignore quantum gravity effects whenever $|H(\tilde{t})|^{-1}\gg \ell_{p}$, where $\ell_{p}$ is the Planck length. We see from the plot that the Hubble radius decreases as the critical energy decreases, i.e., ignoring quantum gravity effects becomes a poorer approximation as the classical limit is approached ($\tilde{\rho}_{cr}\to\infty$). In other words, even though the modified Friedmann equation which produces the bounce is derived from the semi-classical limit of a quantum theory of gravity, the classical limit of the resultant spacetime has a region near the bounce where quantum effects of gravity become important and we would hence expect to detect more particles in this region for larger values of $\tilde{\rho}_{cr}$ than for smaller values. This expectation is also evident from Fig.~\subref*{fig:SPA2} where the gravitational field is clearly changing more rapidly for larger $\tilde{\rho}_{cr}$ (green) and hence we expect more particle production for these values. Put simply, a rapidly contracting universe which contracts to a bounce point with a small Hubble radius -- as occurs for high values of $\tilde{\rho}_{cr}$ -- will lead to an increased rate of quantum particle production at early times relative to smaller values of $\tilde{\rho}_{cr}$. As we will see, this is precisely what we observe in the detector's response.

 \begin{figure}[htb!]
 \centering
 \includegraphics[width=.7\linewidth]{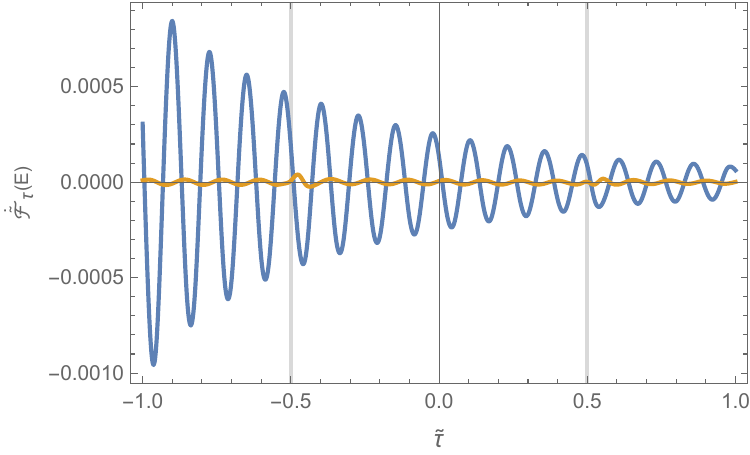}
 \caption{We plot the dimensionless transition rate for the detector passing through the bounce point. The blue curve represents the transition rate when the detector is turned on at $\tilde{\tau}_i=-2$, while the orange curve represents the transition rate when the detector is turned on $\tilde{\tau}_i=-10$. In these plots, we have taken $\tilde{E}=50$, $\tilde{\rho}_{cr}=1$ and the contraction equation of state parameter to be $w=1/3$.}
 \label{fig:transience}
\end{figure}
 One other preliminary issue that must be considered is that of transience, i.e. effects which dominate the signal of the detector during a short time interval after the detector has been turned on (see, for example, Refs.~\cite{Conroy13,Conroy14,Hodgkinson:2012ma,Brenna2016,Hossain:2016klt,LoukoSatz2008}). Hence, we must only take the results as a meaningful signal in the response after transience has subsided. This implies, for example, that if we are interested in understanding the detector's response near the bounce point $t=0$, then it is necessary to turn the detector on sufficiently far into the past of this time. In Fig.~\ref{fig:transience}, we plot the transition rate of two detectors passing through the bounce point, one with initial time $\tilde{\tau}_i=-2$ (blue) and one with $\tilde{\tau}_i=-10$ (orange), all other parameters being identical. What we observe here is that the transient oscillations completely dominate any ``true'' signal in the case where the detector is turned on at $\tilde{\tau}_i=-2$. Hence it is necessary to push the initial time much further into the past. How far into the past depends on the particular parameters one is interested in studying. We find that $\tilde{\tau}_i=-100$ is sufficient for transient effects to be negligible near the bounce for all regions of the parameter space tested.

 	\begin{figure}[htb!]
				\centering
				\resizebox{\linewidth}{!}{%
					\subfloat[\label{fig:BounceLow} $\tilde E=1/2$]{
						\includegraphics[height=6cm]{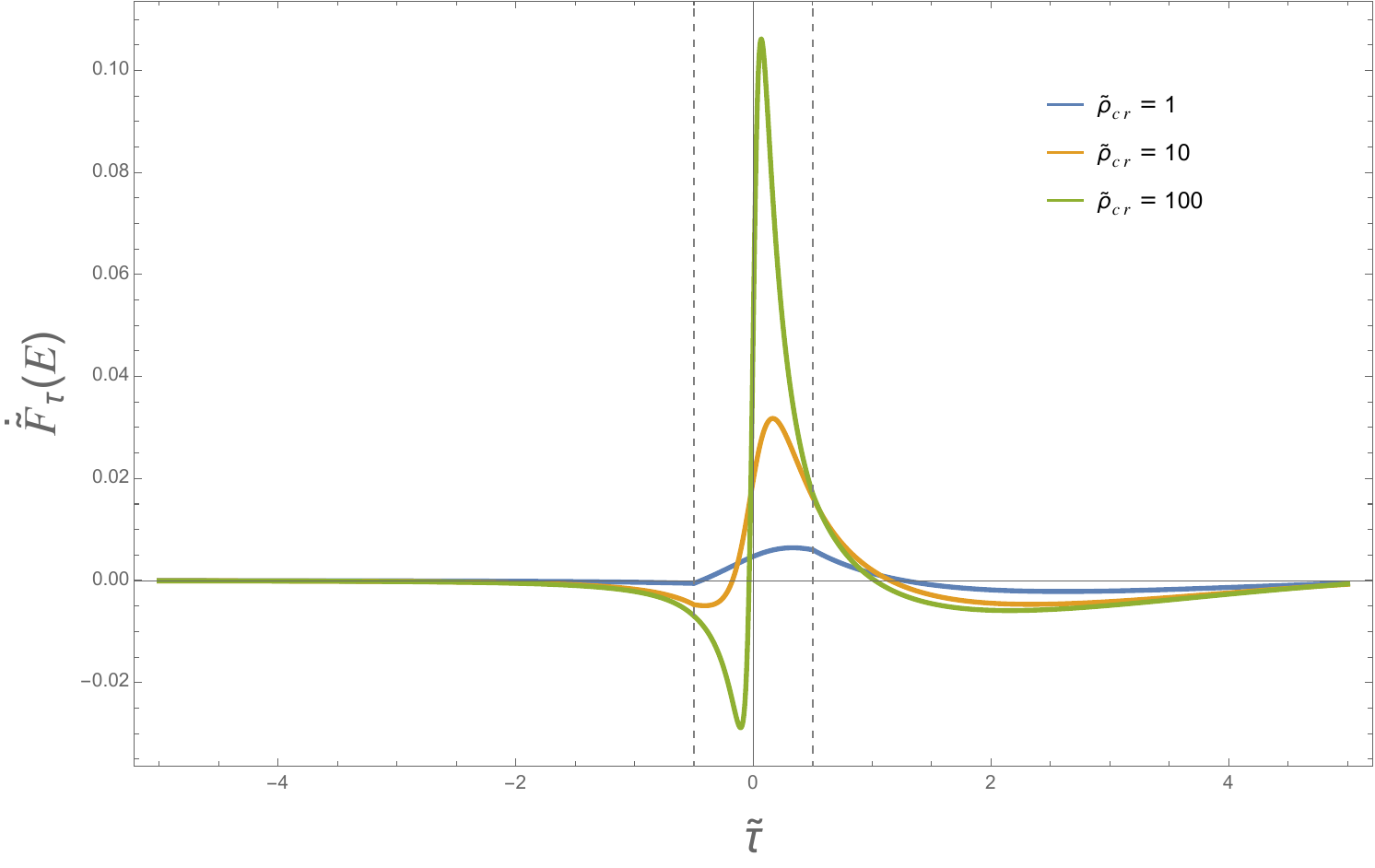}
					}
					\subfloat[\label{fig:BounceHigh}$\tilde E=10$]{
						\includegraphics[height=6cm]{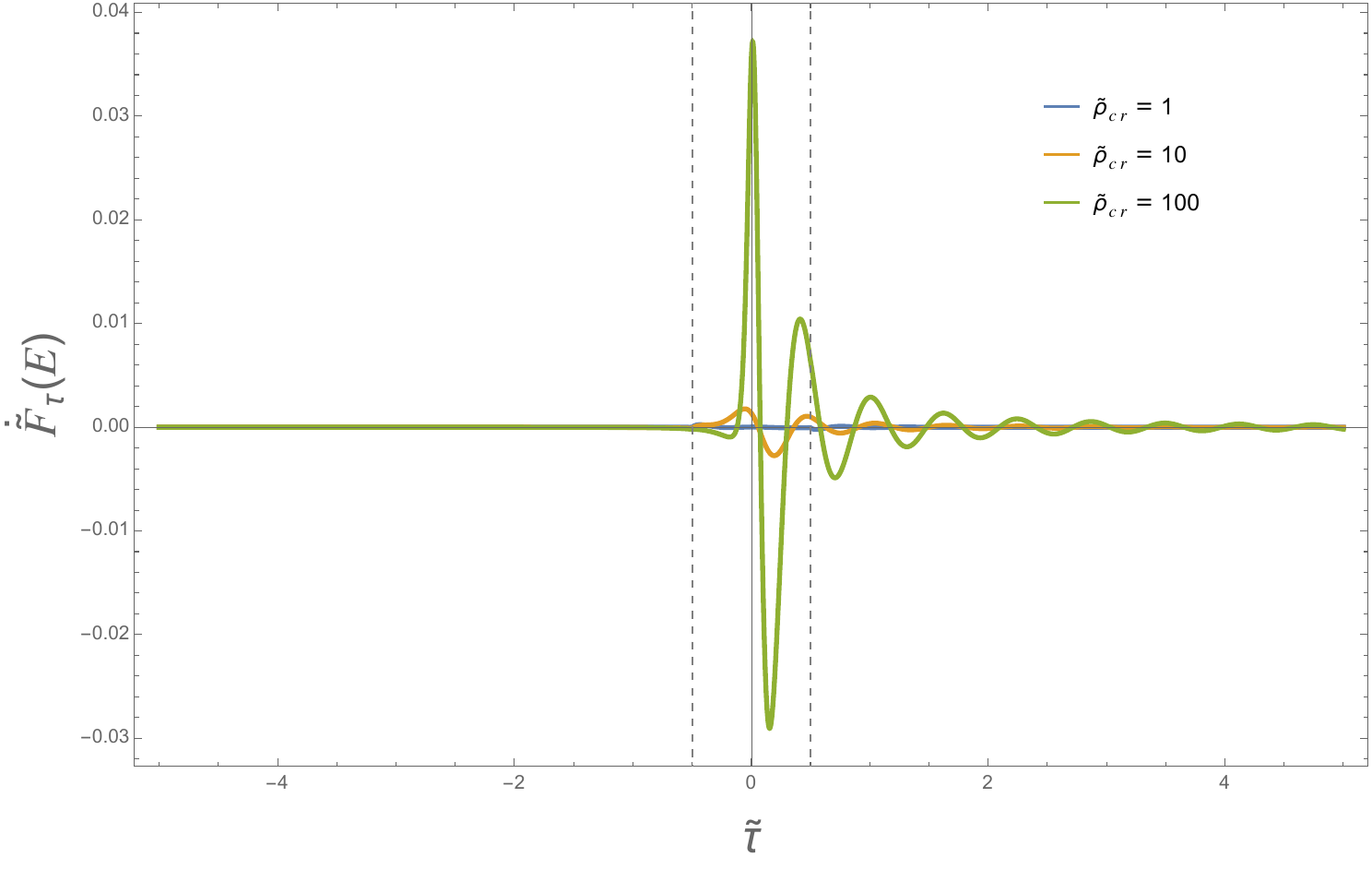}
				}}
				\caption{We plot the transition rates around the bounce for different critical energy densities. The amplitudes of the large oscillations during the bounce are proportional to $\tilde \rho_{cr}$. In each case, we have chosen $\tilde \tau_i=-100$ and $w=1/3$ as the contraction equation of state parameter. The critical energy densities in each plot are $\tilde \rho_{cr}=1$ (blue), $\tilde \rho_{cr}=10$ (orange), and $\tilde \rho_{cr}=100$ (green).}
				\label{fig:BounceRho}
			\end{figure}		
 With the preliminaries out of the way, we look now at how the detector responds as it traverses the bounce. 
 In Fig.~\ref{fig:BounceRho}, we plot the transition rate as a function of time $\tilde t$ and examine how this varies for various choices of the critical energy density. As before, both plots show the transition rate for $\tilde{\rho}_{cr}=1$ (blue curve), $\tilde{\rho}_{cr}=10$ (orange curve) and $\tilde{\rho}_{cr}=100$ (green curve), while the left plot is for $\tilde{E}=1/2$ and the right plot is for $\tilde{E}=10$. Immediately, we see that our earlier intuition was correct in that the larger critical energy densities produces peaks in the transition rate during the bounce phase. What we observe from Fig.~\ref{fig:BounceRho} is a transition rate that is a very slowly varying function of proper time in the contraction phase and then this very quickly changes to a transition rate with large oscillations during the bounce phase. The more precise statement is that the amplitudes of these oscillations is proportional to the critical energy density. The onset of these oscillations in the detector's response is actually non-smooth, but this is simply a feature of our mathematical model which did not possess enough freedom to ensure smoothness of the Hubble parameter. The sharp oscillations during the bounce phase are followed by a series of damped oscillations after the detector emerges from the bounce phase into the matter dominated era. While the amplitudes of the oscillations are proportional to the critical energy density $\tilde{\rho}_{cr}$, the frequency of these oscillations is proportional to the magnitude of the energy gap $|\tilde{E}|$.

\begin{figure}[htb!]
				\centering
				\resizebox{\linewidth}{!}{%
					\subfloat[\label{fig:BounceOmega1} $\tilde \rho_{cr}=1$]{
						\includegraphics[height=6cm]{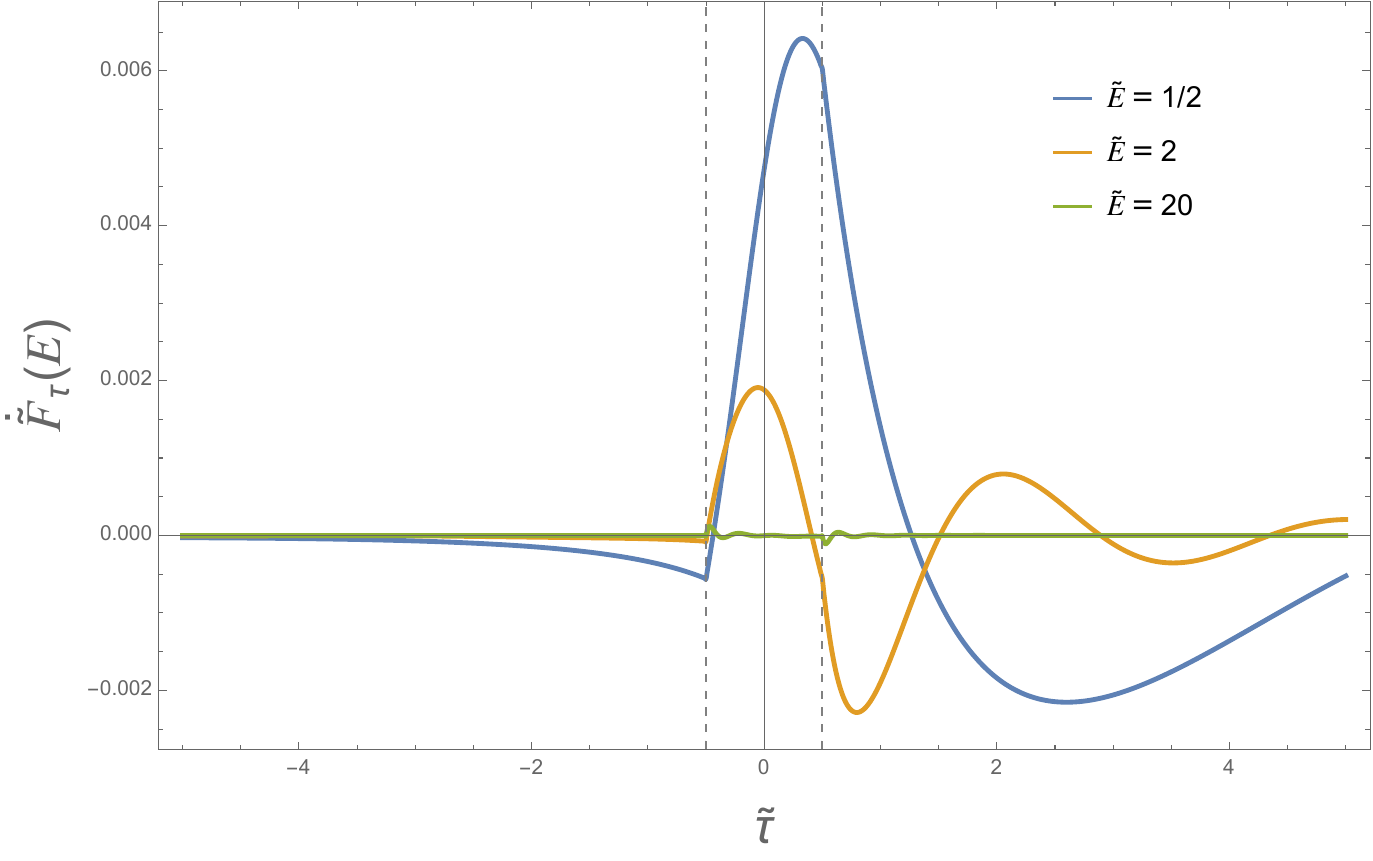}
					}
					\subfloat[\label{fig:BounceOmega2}$\tilde \rho_{cr}=100$]{
						\includegraphics[height=6cm]{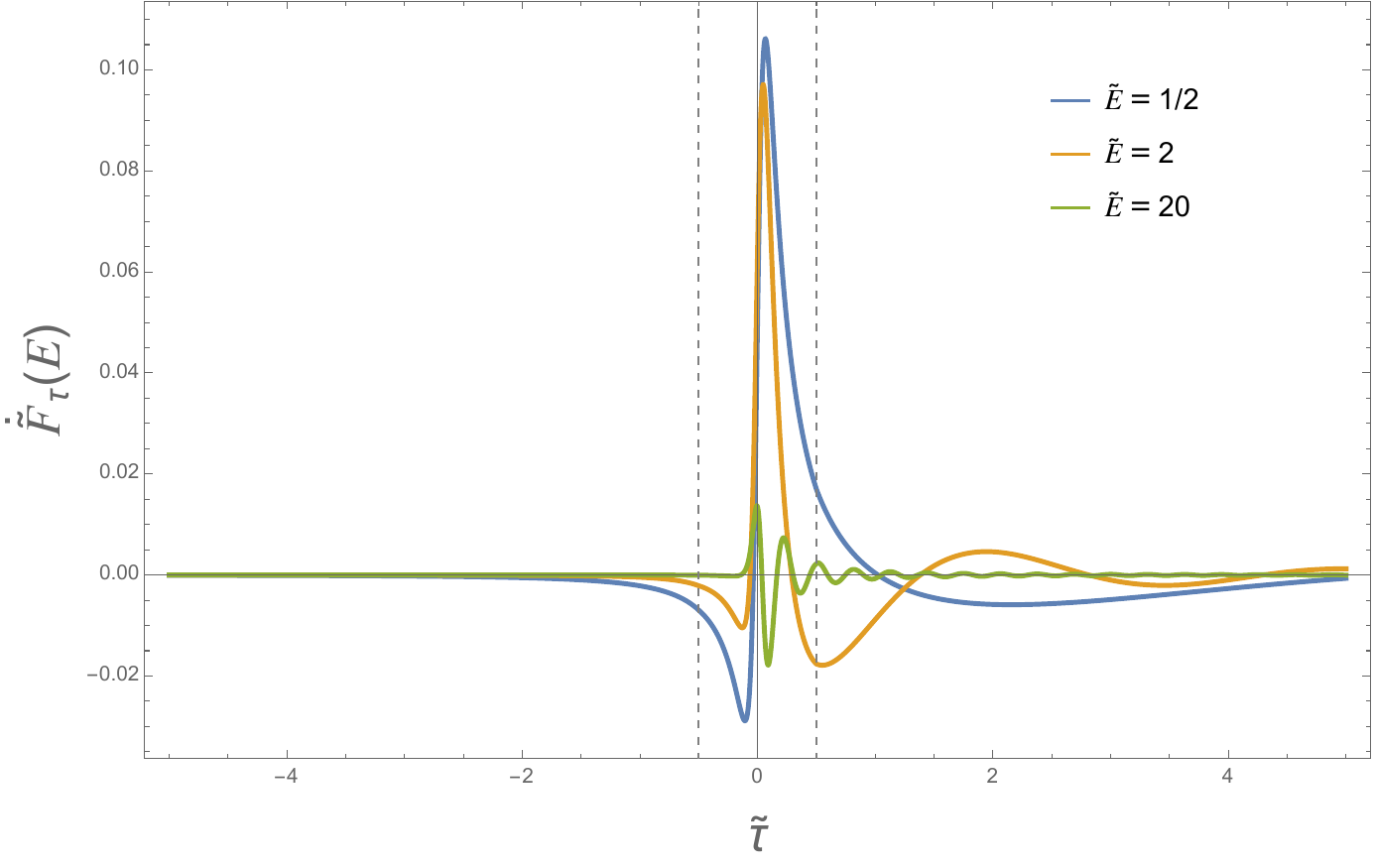}
				}}
				\caption{We plot the transition rate for a detector propagating through the bounce phase for a selection of different energy gaps $\tilde E=1/2$ (blue), $\tilde E=2$ (orange), and $\tilde E=20$ (green). As before, $\tilde \tau_i=-100$ and $w=1/3$ in the contraction phase. The frequency of oscillations in the transition rate is proportional to the energy gap.}
				\label{fig:BounceOmega}
			\end{figure}
\begin{figure}[htb!]
			\centering
		\includegraphics[width=10.7cm]{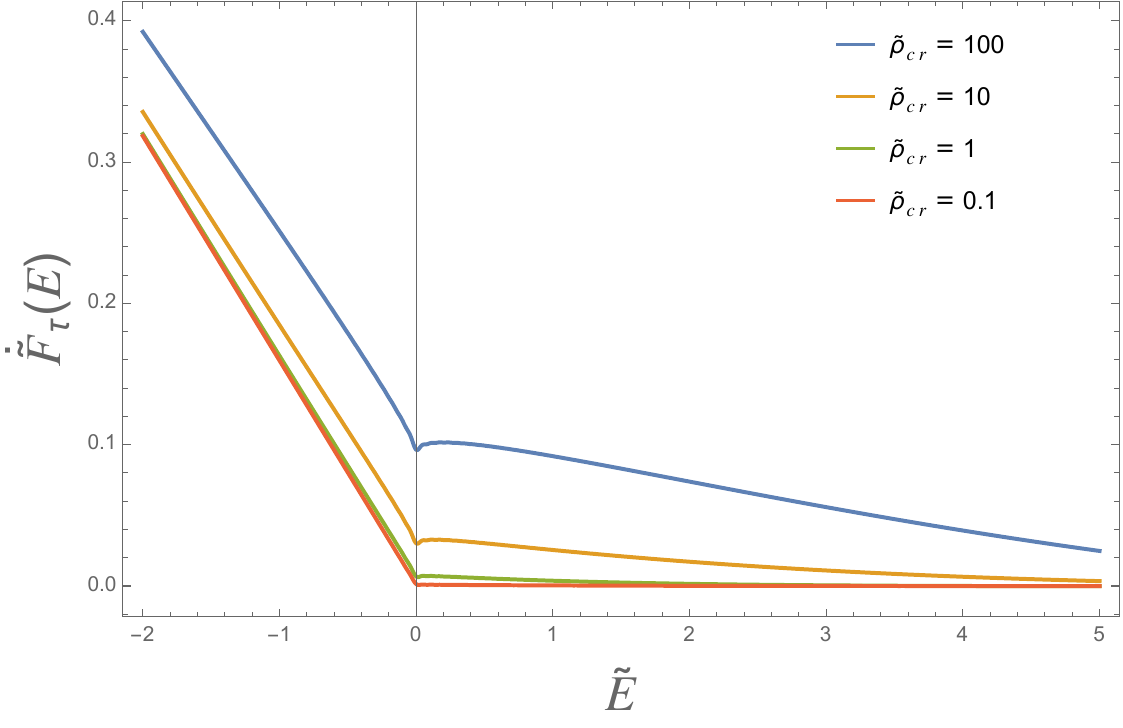}
			\caption{We plot the transition rate for a detector as a function of energy gap at a fixed time near the bounce for a range of critical energy densities.}
			\label{fig:energygapvarrho}
	\end{figure}

 
It is instructive to plot the transition rate for different energy gaps on the same plot for a fixed critical energy density to better ascertain the dependence on the energy gap. 
We plot this in Fig.~\ref{fig:BounceOmega} for a selection of energy gaps with $\tilde E = 1/2$ (blue), $\tilde E = 2$ (orange), and $\tilde E = 20$ (green). We examine these both in the high critical density regime (left) and the low critical density regime (right). It is evident from these plots that the peak in transition rate in the bounce phase is inversely proportional to the energy gap, a result we would expect since it is clearly more probable to excite an electron to a higher energy level when the energy gap is small. Hence the transition rates at low energy gaps will have oscillations with higher amplitudes. What is interesting, however, is that the dependence of the transition rate on positive energy gaps appears to be quite sensitive to the critical energy density. We see this in Fig.~\ref{fig:energygapvarrho}, where the transition rate as a function of $\tilde{E}>0$ at a given time in the bounce phase is almost constant for small critical energy densities but not for larger critical energy densities. As shown in Appendix \ref{sec:Limit}, the transition rate asymptotes to $|\tilde{E}|/2\pi$ for large negative energy gap while it tends to zero for large positive energy gap. These features can also be seen in the graph in Fig.~\ref{fig:energygapvarrho}.

As already mentioned, we note a clear lack of smoothness in the transition rate at the transition between eras. The transitions in and out of the bounce phase occur at $\tilde \tau=\pm 1/2$ in all our plots. The non-smoothness is most evident in Fig.~\subref*{fig:BounceOmega1}. The reason is that our analytic model contains  enough free parameters to make the scale factor $a$ and conformal time trajectory $\eta$ smooth, but there is not enough freedom in the integration constants to make $\dot{a}(\tau)$ and hence the Hubble parameter smooth. While the integrand in the transition rate (\ref{eq:FDotTilde}) doesn't explicitly involve derivatives of the scale factor, cancellation of the singularities at $s=0$ does involve a Taylor expansion which involves the derivative. Hence the transition rate is not smooth on the boundary between eras. However, when we take the critical energy density to be large, then the transition rate passing from the bounce phase to matter domination is approximately smooth. If we also take the contraction phase equation of state parameter to be $w=1/3$, then the transition rate as the detector crosses into the bounce phase from the contraction phase is also approximately smooth. This is because taking $w=1/3$ in the contraction phase and $\tilde{\rho}_{cr}$ large in the bounce phase implies we are solving (approximately) the same standard Friedmann equation for radiation domination in both cases. Unfortunately, for small values of $\tilde{\rho}_{cr}$, this non-smoothness is responsible for some artefact in the detector. One might hope that, even if the transition between eras was smooth but very rapid, then this analytical model is still capturing the essential features.

  \subsection{Relics of pre-bounce physics}
  \label{sec:EOS}
    \begin{figure}[htb!]
			\centering
			\resizebox{\linewidth}{!}{%
				\subfloat[\label{fig:SPB1} Bounce phase throat]{
					\includegraphics[height=4.7cm]{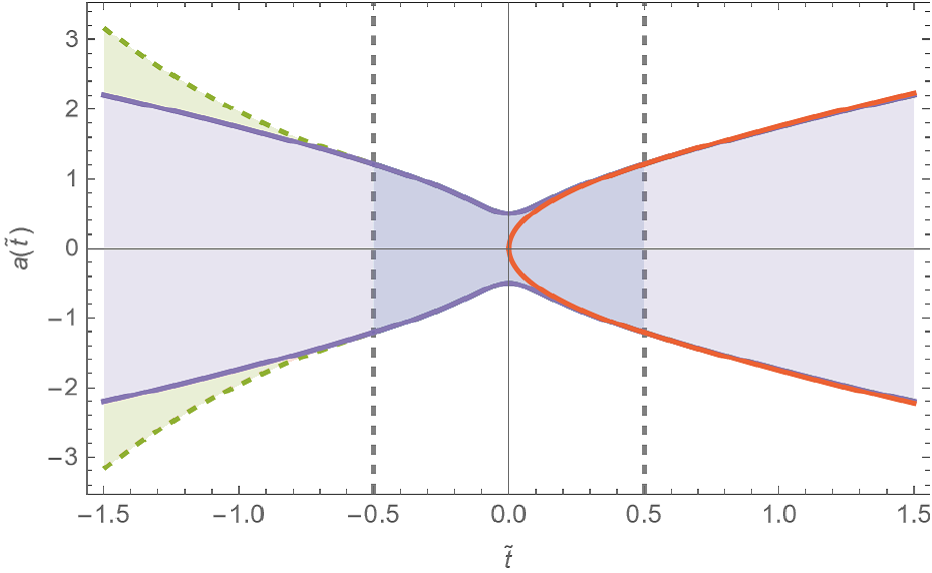}
				}
				\subfloat[\label{fig:SPB2}Scale factor at early times]{
					\includegraphics[height=4.7cm]{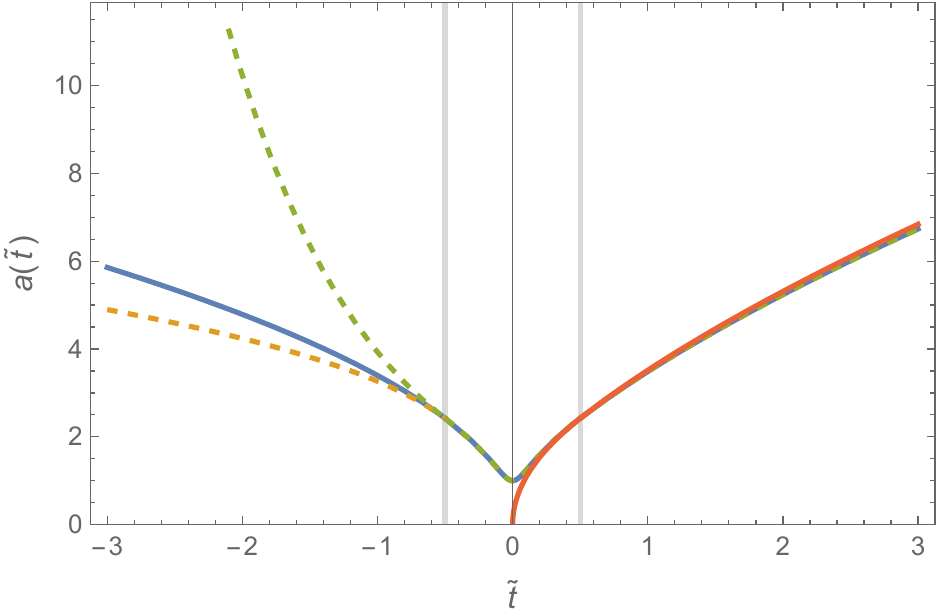}	
			}}
			\\	\resizebox{\linewidth}{!}{
				\subfloat[\label{fig:SPB3} Hubble parameter at early times]{
					\includegraphics[height=4.7cm]{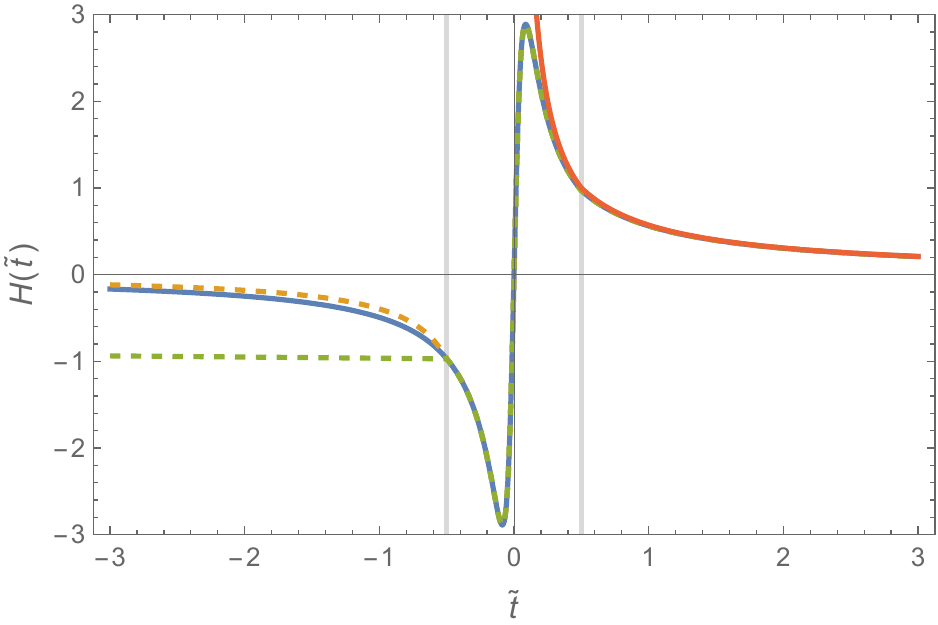}
				}
				\subfloat[\label{fig:SPB4} Hubble radius $1/|\tilde H(\tilde t)|$]{
					\includegraphics[height=4.7cm]{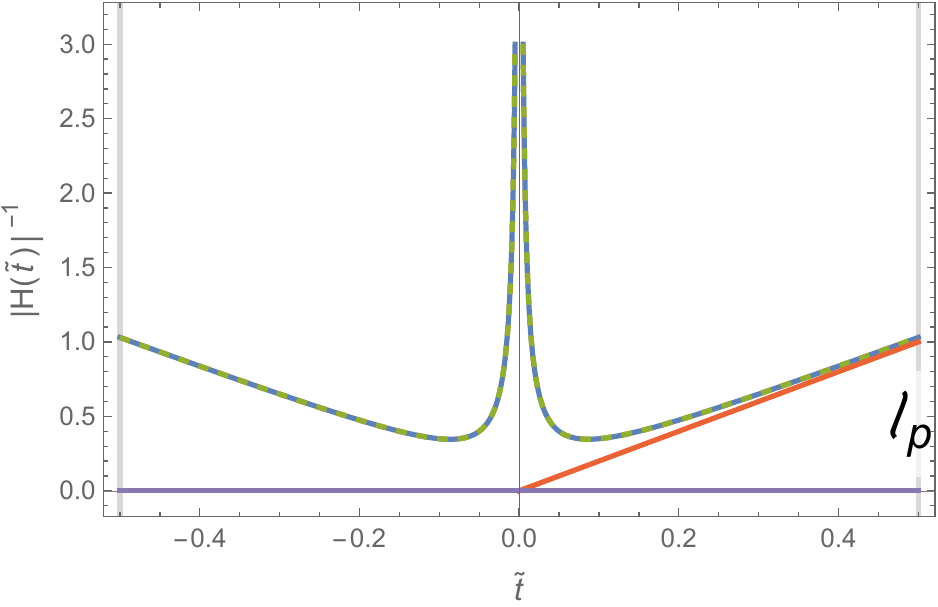}
									}}
				 \caption{We illustrate the bounce phase throat as the blue shaded region in Fig.~\protect\subref*{fig:SPB1}. There are two matter-dominated regions (purple) with $w=0$, one pre-bounce and one post-bounce. The green curve is for a contraction phase with equation of state $w=-1$.
     In addition, we include a singular \emph{big bang} model in red which has been tuned to agree with the  non-singular models at $\tilde t=\tilde t_m$ where, for these plots, we have chosen $\tilde \rho_{cr}=100$. In Figs.~\protect\subref*{fig:SPB2}--\protect\subref*{fig:SPB4}, we present the scale factor, (dimensionless) Hubble parameter, and (dimensionless) Hubble radius for a variety of contraction phase equation of state parameters where  $w=1/3$ (blue), $w=1$ (orange), and $w\to-1$ (green).
     }
    \label{fig:SpacetimeB}
			\end{figure}
   
In this subsection, we investigate the dependence of the transition rate on the equation of state parameter $w\ge -1$ during the contraction phase. In particular, we would like to establish to what extent the particle detector ``remembers'' the contraction phase after it has passed through the bounce and whether we can distinguish between different pre-bounce equations of state through an analysis of the particle rate at late times. 

The choices under consideration here are (i) a radiation dominated contraction phase with $w = 1/3$; (ii) a contraction phase with $w=1$ as proposed by Steinhardt and Ijjas in Ref.~\cite{Ijjas:2018qbo}, since the anisotropy factor scales as $\Omega_{a}\sim a^{3(w-1)}$, this factor is therefore constant as the bounce is approached preventing the anisotropy from growing and triggering mixmaster chaotic behaviour; and (iii) the $\Lambda$CDM bounce scenario which supposes some cold dark matter in the contraction phase whereby $-1\leq w\leq0$ (see Refs.~\cite{Cai:2014jla,Brandenberger:2016vhg}). In our investigations we take a $\Lambda$CDM model with $w \to-1$. We plot the scale factor, Hubble parameter and Hubble radius for these different contraction equations of state in Fig.~\ref{fig:SpacetimeB}. We also include in these plots a singular, classical Big Bang model where the scale factor has been normalized to agree with the scale factor in a $\tilde{\rho}_{cr}=100$ bouncing cosmology  at $\tilde t=\tilde t_m$ (later we normalize the scale factor in the Big Bang model to agree with the $\tilde{\rho}_{cr}=1$ bouncing cosmology at $\tilde{t}=\tilde{t}_{m}$ but the transition rate does not depend on this choice of normalization).
	
        \begin{figure}[htb!]
				\centering
				\resizebox{\linewidth}{!}{%
					\subfloat[\label{fig:InitialTime1} $\tilde{E}=0.1$, $\tilde{\rho}_{cr}=0.1$]{
						\includegraphics[width=8.2cm]{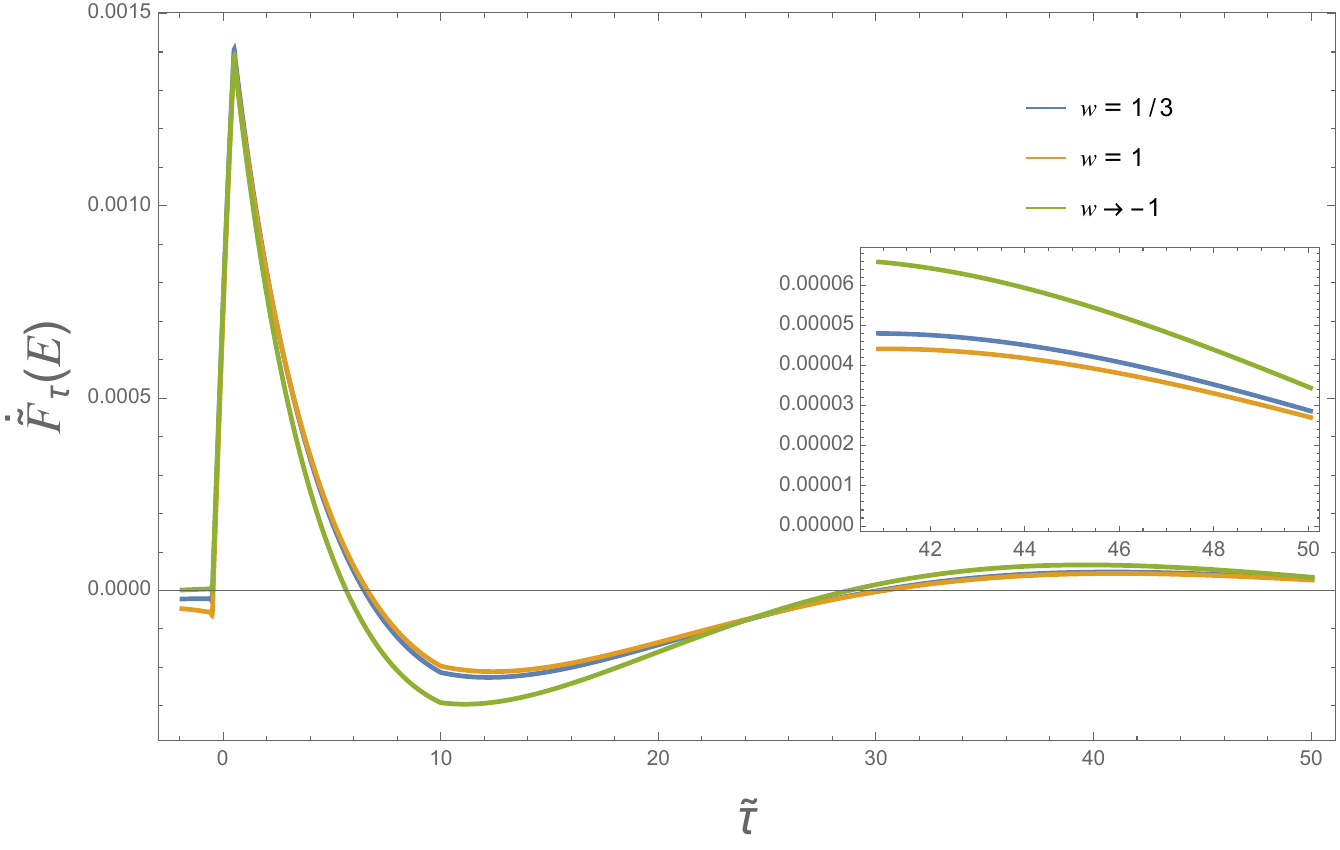}
					}
					\subfloat[\label{fig:InitialTime2}$\tilde{E}=0.1$, $\tilde{\rho}_{cr}=10$ ]{ 
						\includegraphics[width=8.2cm]{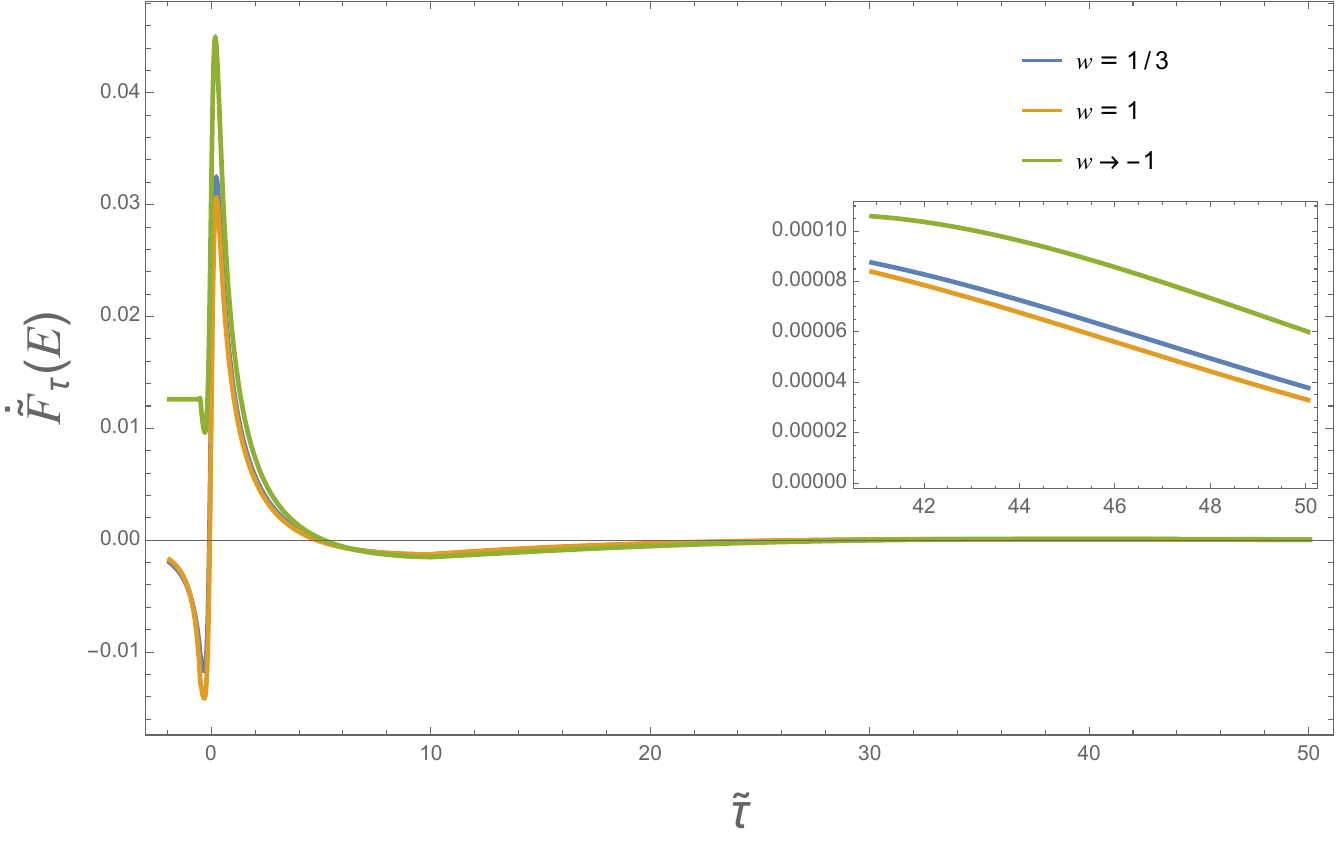}	
				}}
						\caption{We plot the transition rate as a function of the detector's proper time for a selection of contraction phase equation of state parameters. In both plots above we take the initial time to be $\tilde{\tau}_{i}=-100$ and the energy gap to be $\tilde{E}=0.1$. Fig.~\protect\subref*{fig:InitialTime1} has $\tilde{\rho}_{cr}=0.1$ while  Fig.~\protect\subref*{fig:InitialTime2} has $\tilde{\rho}_{cr}=10$. } 
				\label{fig:PostEOS}
			\end{figure}
     \begin{figure}[htb!]
				\centering
			\resizebox{\linewidth}{!}{
					\subfloat[\label{fig:HighELowRho} $\tilde{E}=5$, $\tilde{\rho}_{cr}=0.1$]{
						\includegraphics[width=8cm]{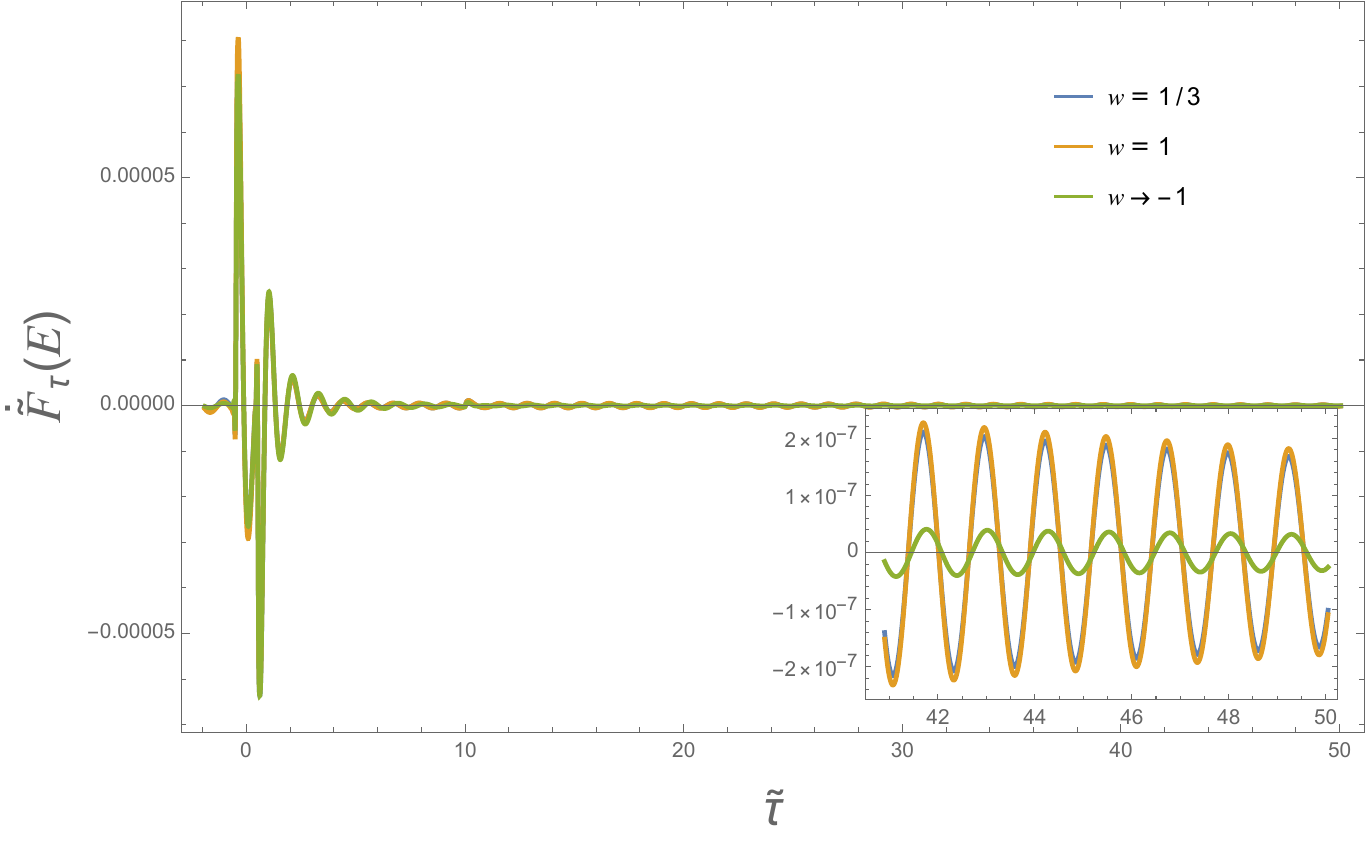}
					}
					\subfloat[\label{fig:HighEHighRho} $\tilde{E}=5$, $\tilde{\rho}_{cr}=10$ ]{
						\includegraphics[width=8cm]{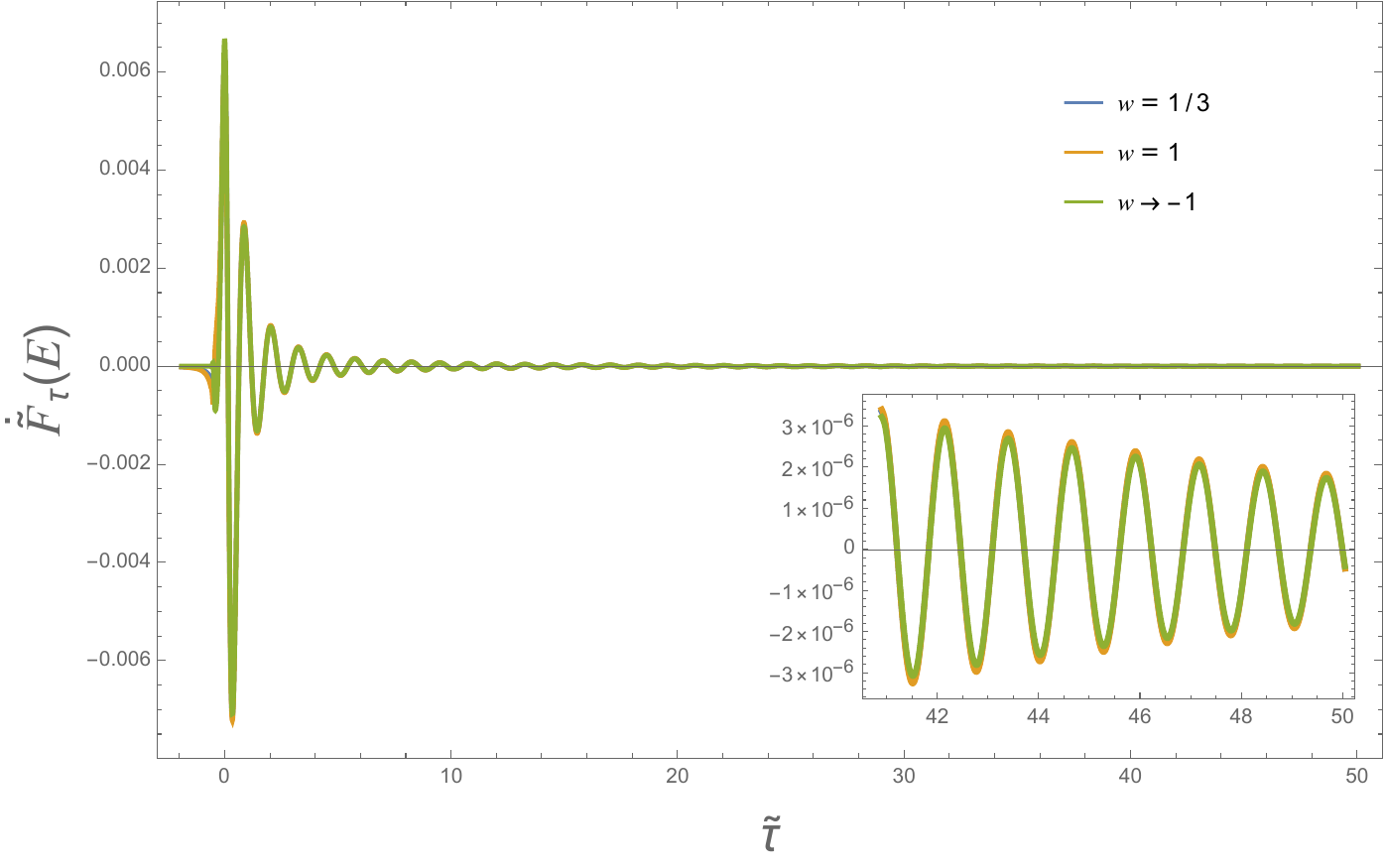}	
				}}
				\caption{We plot the transition rate as a function of the detector's proper time for a selection of contraction phase equation of state parameters. In both plots we have turned the detector on at $\tilde{\tau}_{i}=-100$ and taken a reasonably large energy gap of $\tilde{E}=5$. In Fig.~\protect\subref*{fig:HighELowRho}, we have $\tilde{\rho}_{cr}=0.1$ while  Fig.~\protect\subref*{fig:HighEHighRho} has $\tilde{\rho}_{cr}=10$}
				\label{fig:HighE}
			\end{figure}

We now turn our attention  to the question at hand, namely, we ask in what region of the parameter space is the detector's response in the post-bounce phase most sensitive to different choices of equation of state $w$ in the contraction phase. That is, can some remnant of pre-bounce physics survive in the transition rate at later times? In Figs.~\subref*{fig:InitialTime1} and~\subref*{fig:InitialTime2}, we plot the transition rate for different choices of equation of state for a very small energy gap. The left-hand plot is for $\tilde{\rho}_{cr}=0.1$ where we would expect quantum gravity effects to be important while the right-hand plot is for $\tilde{\rho}_{cr}=10$. Nevertheless, the sensitivity to the equation of state parameter seems to be generally the same in either case, we observe an inverse proportionality between the equation of state parameter $w$ and the transition rate at late times. In other words, the $w=-1$ model has the highest transition rate while the $w=1$ model has the lowest transition rate. The $w=-1$ model appears to have a significantly higher transition rate than any model with positive $w$.

   \begin{figure}[htb!]
				\centering
			\resizebox{\linewidth}{!}{
					\subfloat[\label{fig:Later} $\tilde E=1$, $\tilde{\rho}_{cr}=0.08$]{
						\includegraphics[width=8cm]{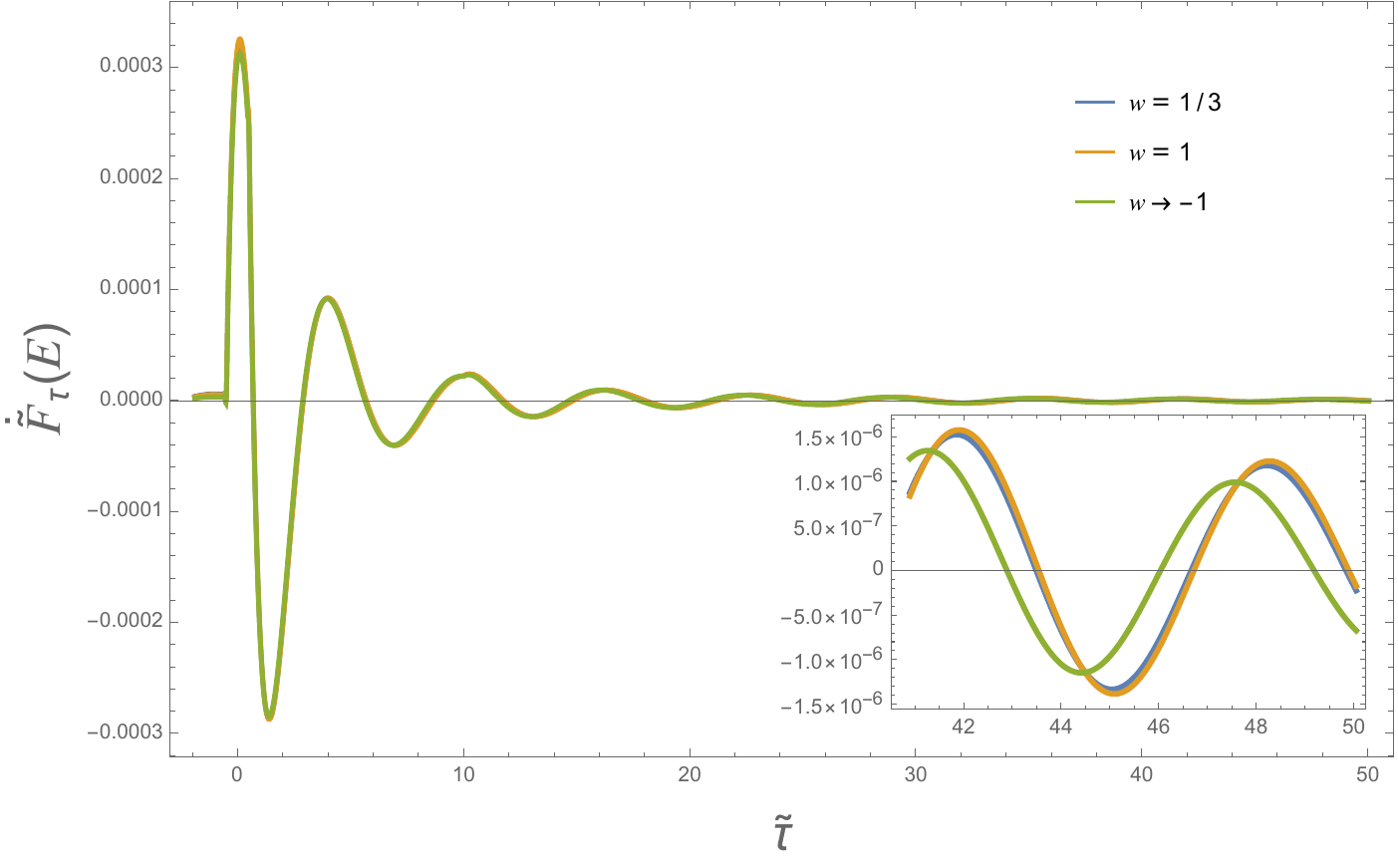}
					}
					\subfloat[\label{fig:EOSTodayb} $\tilde{E}=1$, $\tilde{\rho}_{cr}=0.02$ ]{
						\includegraphics[width=8cm]{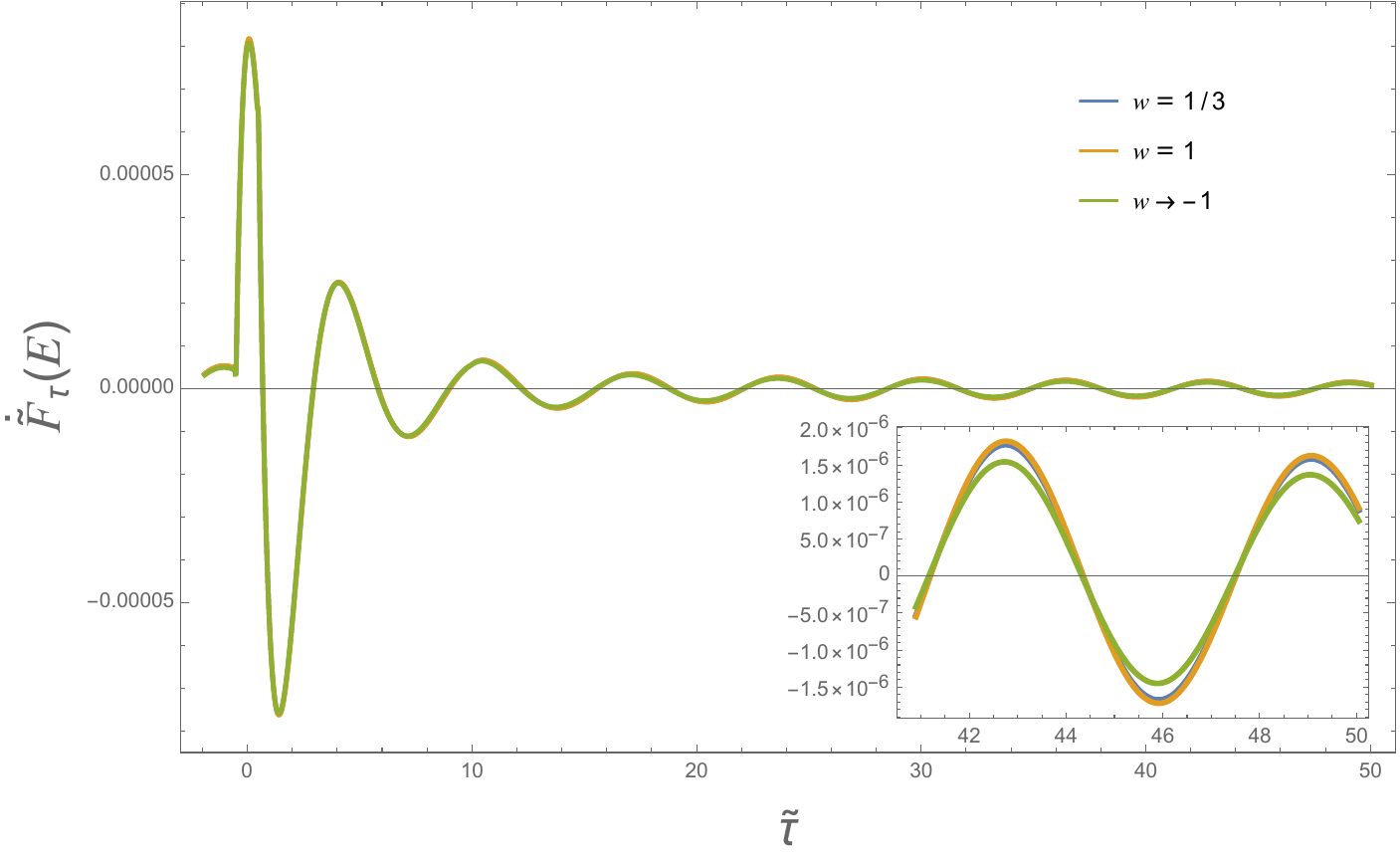}	
				}}
				\caption{We again examine the late-time transition rates for different equations of state, deep in the quantum bounce regime. In each plot, the detector is turned on at $\tilde{\tau}=-100$ and the energy gap is $\tilde{E}=1$. On the left, we illustrate that there is a very small region of the parameter space in the small $\tilde{\rho}_{cr}$ regime where the phase of the oscillations for the $w=-1$ equation of state are out of sync with the oscillations for the other equation of state parameters. We see this in the inset plot on the left. Changing the critical energy density by even a small amount, as in the plot on the right, and the oscillations are back in phase with one another.} 
				\label{fig:EOSToday}
			\end{figure}
For larger positive energy gap, the analysis is not as straightforward since the transition rate is a rapidly oscillating function of time for large $\tilde{E}$. For large energy gap and large critical energy density, there is very little disparity in the transition rate at late times for the different equations of state. This can be seen in Fig.~\subref*{fig:HighEHighRho} where the plots are very difficult to distinguish. However, for large energy gap and small critical energy density, we see from the inset plot in Fig.~\subref*{fig:HighELowRho} a clear distinction between the $w=-1$ model and the other models, where the amplitudes of oscillations in the transition rate are significantly suppressed at late times for the $w=-1$ case relative to $w=1/3$ and $w=1$. The phase of the late-time oscillations also appear to be approximately equal for different equation of state parameters, except that we found numerically that there is a very small region of parameter space with small $\tilde{\rho}_{cr}$ where these oscillations for $w=-1$ are out of phase relative to the oscillations for other contraction phase equations of state. This can be seen in Fig.~\ref{fig:EOSToday} where we plot transition rates for different contraction phase equations of state with $\tilde{\rho}_{cr}=0.08$ and it is clear that the $w=-1$ (green curve) oscillations are out of phase relative to the other plots. However, if we change the critical energy density by even a small amount, the oscillations are back in phase for all equations of state, as can be seen in the right plot in Fig.~\ref{fig:EOSToday} which has $\tilde{\rho}_{cr}=0.02$. 
			
				\subsection{Signatures of non-singular cosmologies}		
    \label{sec:BB}
    Finally, in this subsection, we explore the question of whether our particle detector can distinguish between a singular \emph{big bang} cosmology and a non-singular bouncing cosmology.  We consider bouncing cosmologies with a range of critical energy densities keeping in mind the limit $\tilde \rho_{cr}\to\infty$ retrieves the classical Friedmann equation. We would therefore expect the difference in the detector's response to be most sensitive to the small $\tilde{\rho}_{cr}$ bouncing cosmology. As the detector is sensitive to the detection time, we demand that both detectors begin detecting at the same initial time $t_{i}>0$ with $t_{i}<<1$. Of course this is not a realistic scenario since ideally we would like to turn our detectors on at very late times to see if a detector today has a long memory of the past. However, our simple analytical model does not encode any difference between these two scenarios if the detector is turned on at late times. This would require a more sophisticated numerical model for the spacetime geometry that doesn't simply paste the dominant scale factors together.
    
\begin{figure}[htb!]
				\centering
				\resizebox{\linewidth}{!}{%
					\subfloat[\label{fig:Sig1}  Early CMB formation ($\tilde E=2$)]{
						\includegraphics[height=4.7cm]{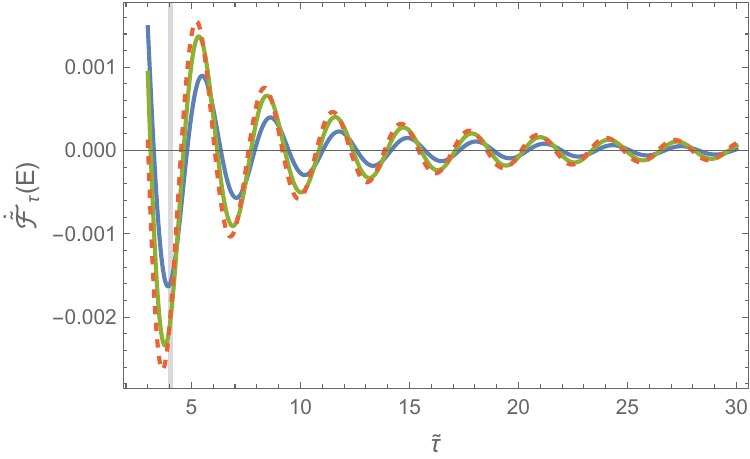}
					}
					\subfloat[\label{fig:Sig2} Early CMB formation ($\tilde E=200$)]{
						\includegraphics[height=4.7cm]{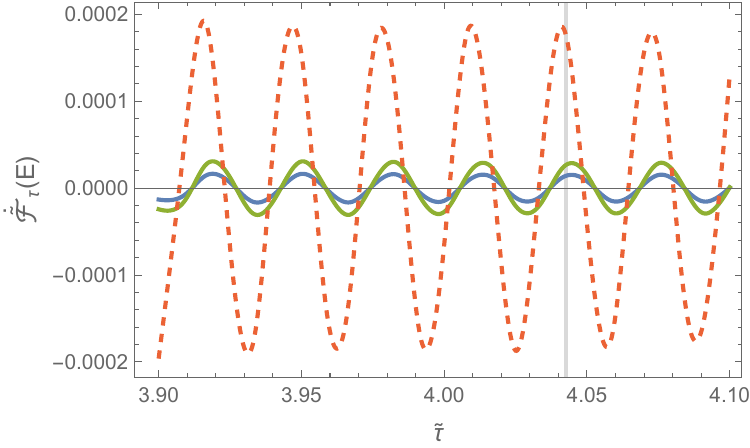}
				}}	\\
					\resizebox{\linewidth}{!}{%
					\subfloat[\label{fig:Sig3} Matter-dominated era ($\tilde E=2$)]{
						\includegraphics[height=4.7cm]{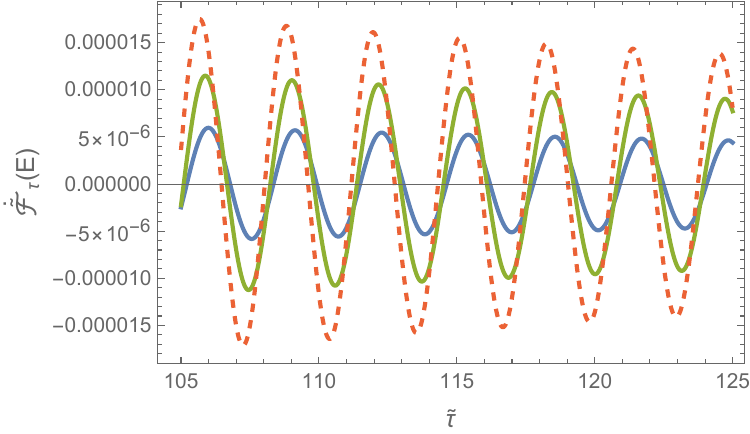}
					}
					\subfloat[\label{fig:Sig4} Matter-dominated era ($\tilde E=200$)]{
					\includegraphics[height=4.7cm]{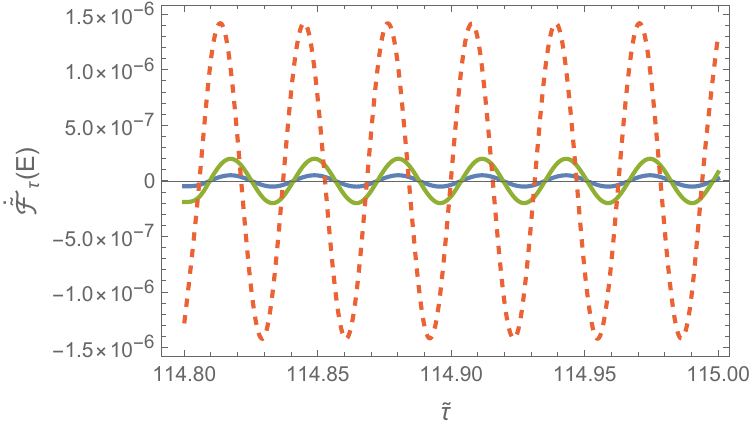}	
				}}
						\caption{ We compare the transition rate of a detector in a bouncing cosmology with $\tilde \rho_{cr}=1$ (blue) and $\tilde \rho_{cr}=100$ (green) with a classical, singular Big Bang model in red. Both detectors begin detecting just after $\tilde t=0$.}
				\label{fig:BB}
			\end{figure}

    To normalize the scale factor in the singular model, we demand that the scale factor for the singular cosmology agrees with the scale factor for the $\tilde{\rho}_{cr}=1$ non-singular cosmology at $\tilde t=\tilde t_m$, which is the start of the matter-dominated era and is governed by the standard Friedmann equation. We note that the transition rate itself does not depend on how the scale factor has been normalized. For times $\tilde{t}>\tilde{t}_{m}$, the singular and non-singular cosmologies are identical and as such, the detector will only have a brief window on the interval $0<\tilde{t}_{i}< \tilde t<\tilde t_m$ (where $\tilde t_m=1/2$ in our dimensionless time) in which the detector is sensitive to whether the universe has a bounce or a spacetime singularity at $t=0$. We intend to investigate what region of the parameter space is the detector most sensitive to this dichotomy and if this is apparent in the detector's response even at late times. For context, the time which corresponds to approximately `today' in our model is $\tilde t\approx 146,809$. To this end, in Fig.~\ref{fig:BB}, we compare the transition rate of such detectors at the time of the early formation of the CMB  ($\tilde \tau\approx 4.04$) and at a later time, deep within the matter-dominated era which corresponds to approximately $10.8$ million years after $t=0$. For each time-frame, we consider both the high and low energy gap regimes. As in the previous section, we find that the frequency of the oscillations is proportional to the energy gap and, moreover, we find that the high energy gap regime better highlights disparities between the theories. While from Figs.~\subref*{fig:Sig1} and~\subref*{fig:Sig3}, we see that the transition rate of a detector propagating in a non-singular spacetime is generically more suppressed than in the singular case, this pattern is more pronounced as we move across to the high energy gap regime in Fig.~\subref*{fig:Sig2} and~\subref*{fig:Sig4}. We also note, as expected, that the transition rate for a detector in a bouncing cosmology with a smaller Hubble radius (see green curve) and larger $\tilde{\rho}_{cr}$ approaches the transition rate for a detector in a singular cosmology (red), however even for bouncing cosmologies which have a very small Hubble radius near the bounce, in the high energy gap regime the amplitudes of the transition rate are still suppressed by an order of magnitude relative to the \emph{big bang} model. 
   
\begin{figure}[htb!]
				\centering
				\resizebox{\linewidth}{!}{%
					\subfloat[\label{fig:today1}  Today ($\tilde E=2$)]{
						\includegraphics[height=4.7cm]{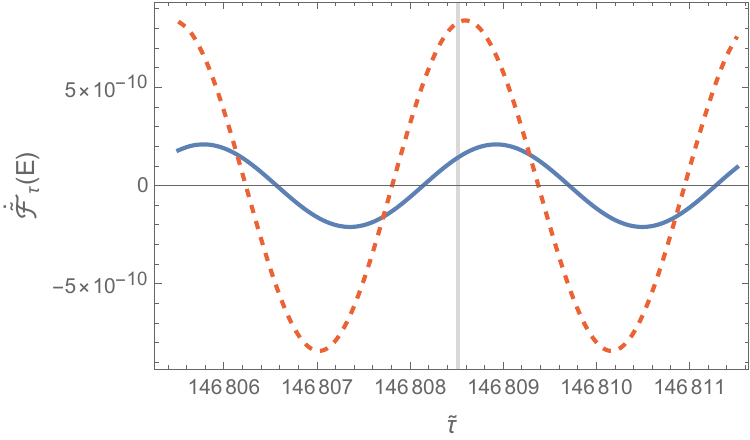}
					}
					\subfloat[\label{fig:today2} Today ($\tilde E=200$)]{
						\includegraphics[height=4.7cm]{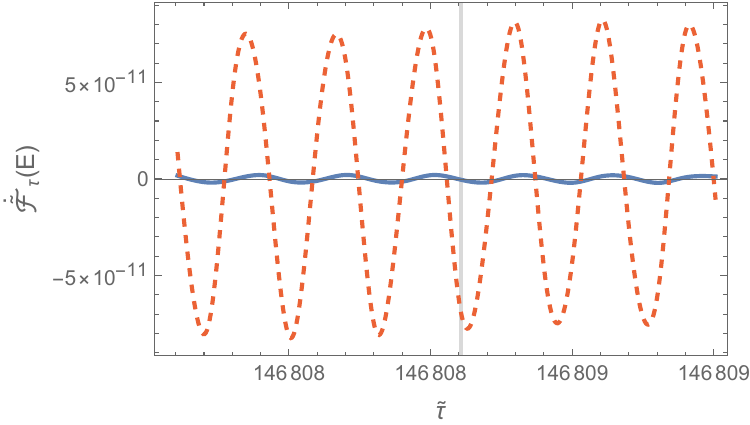}
				}}
						\caption{ We compare the transition rate of a detector in a bouncing cosmology with $\tilde \rho_{cr}=1$ (blue) with a classical, singular Big Bang model in red. The transition rate has been evaluated at what corresponds to approximately `today' in our model. Both detectors begin detecting just after $\tilde t=0$. }
				\label{fig:BB2}
			\end{figure}

   We further note from Fig.~\ref{fig:BB2} that this behaviour persists deep into the late time dark energy phase at a time which corresponds to `today' in our model. While the amplitudes of the transition rate for a detector in both the singular and bouncing cosmologies are very small, the relative suppression of the bouncing model is still present.

			\section{Conclusions}
			In this paper, we investigated how a study of semi-classical particle production can shed light on early universe physics. Working within the Unruh-DeWitt particle detector model in the QFTCS semiclassical approximation, we studied the transition rate of a detector as it travelled through a non-singular bouncing universe.
   
   In order to render the calculation of the transition rate tractable, we employed an analytic model for the spacetime of the bouncing cosmology. In all cosmological eras apart from the bounce phase, we employed the standard Friedmann equation, assuming only the dominant form of the energy density in that era. In the contraction phase, we consider a general constant equation of state parameter and investigate how different choices manifest in the detector's response. For the bounce phase, we assumed a modified Friedmann equation inspired by the semi-classical limit of a \emph{loop quantum cosmology} model. Regardless of the modification's origin, we take a phenomenological philosophy and take this simply to be the governing equation untethered to any particular underlying quantum gravity or modified gravity model. The scale factors for these different cosmological eras are then glued together enforcing smoothness at the transitions.
   
The advantage of this analytical model is that it facilitates a reasonably straightforward construction of the Wightman two-point function, which is the essential ingredient in the computation of the transition rate. Since we assume a flat FLRW spacetime and that the quantum field coupled to the detector is conformally coupled to the background geometry, the Wightman function is conformal to the Minkowski two-point function in conformal coordinates, where the conformal factor is simply the scale factor. In principle, we could construct the Wightman function without assuming the simple analytical approximation adopted here but the scale factor and the conformal time would have to be solved numerically. Because of the computational cost, this would also limit the region of parameter space that could be explored. Nevertheless, we hope to return to a numerical model in the future.

Armed with a simple analytical model for the background spacetime and a readily calculable Wightman function, the transition rate for the entire parameter space was open for exploration. The parameter space is rather large, for example, we have the detection time, the energy gap, the critical energy density of the correction in the bounce phase and the equation of state parameter in the contraction phase. We wanted to explore this parameter space with three principal questions in mind: (i) Does the transition rate of a detector exhibit any robust general features as it traverses the bounce? (ii) Is the equation of state parameter for the contraction phase imprinted on the transition rate at late times? and (iii) Are Big Bang cosmologies distinguishable from bouncing cosmologies in the transition rate at late times?

The answers to the questions above do indeed turn out to be very sensitive to what region of the parameter space we are in. Nevertheless, some robust features do emerge. We found that the particle detector always responds sharply to the bounce phase with the transition rate undergoing a transition from a very slowly varying function of time during the contraction phase to a series of large oscillations during the bounce phase. The amplitude of these oscillations is inversely proportional to the critical energy density while their frequency is proportional to the energy gap. The residue of these oscillations persists throughout the rest of the detector's trajectory, although the oscillations are damped with time. We found further that, for different contraction phase equation of state parameters, the differences in these late-time damped oscillations are typically very small and hard to distinguish. However, there are some small regions of the parameter space where the disparity is amplified. For example, for high energy gaps and small critical energy density, the oscillations in the transition rate for the $w=-1$ equation of state is suppressed significantly relative to other equations of state. Interestingly, this region of parameter space also appears to be the `goldilocks' zone for distinguishing between Big Bang cosmologies and non-singular bouncing cosmologies at very late times. That is, when the energy gap is large, the residual oscillations in the transition rate at early times is significantly more suppressed at late times for bouncing cosmologies with smaller critical energy density than compared with the singular Big Bang cosmology.

Since this study involves only the conformally invariant field and the metric is conformal to Minkowski spacetime, it is difficult to disentangle the true effect of particle production due to the dynamical nature of the spacetime. In order to distil the effects of the changing background geometry -- inasmuch as it is possible within the particle detector framework -- one needs to break conformal invariance and consider alternative field couplings. Computing the transition rate for non-conformal coupling constants is technically more challenging since the Wightman function is not known in closed form but only as a mode sum. In this case, the $1/(4\pi^{2}s^{2})$ Hadamard singularity in the integrand of Eq.~(\ref{eq:transitionratesharp}) must be expressed as a mode-sum which can be subtracted mode-by-mode from the Wightman function. Performing the integral in Eq.~(\ref{eq:transitionratesharp}) numerically can be challenging, especially if the mode-sum representation of the regularised Wightman function is slowly convergent. We hope to return to this calculation in the future.


			\section*{Acknowledgements}{A.C. is supported by the Czech Science Foundation grant No. 22-14791S. We would like to thank Dr. Yi-fu Cai and Professor Edward Wilson-Ewing for helpful discussions during the preparation of this article.}

						\appendix

			\section*{Appendix}
				\section{Regularisation and sharp-switching}
	\label{sec:Sharp}

\begin{figure}[htb]
	\centering
	\includegraphics[width=\linewidth]{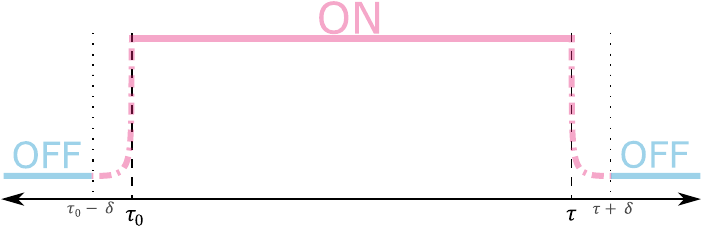}
	\caption{Profile of a smooth switching function.}
  \label{fig:onoff}
\end{figure}

In Ref.~\cite{Satz2007}, the sharp-switching limit of the response function (\ref{responsereg}) was obtained by starting out with a smooth switching function of the form
\begin{equation}
	\label{eq:switchdef}
	\chi(u)=h_{1}\left(\frac{u-\tau_{0}+\delta}{\delta}\right)\times h_{2}\left(\frac{-u+\tau+\delta}{\delta}\right),
\end{equation}
where $\tau>\tau_{0}$, $\delta>0$, and $\tau_{0}$ is some initial proper time. The functions $h_{1}$ and $h_{2}$ are arbitrary smooth functions satisfying
\begin{align}
	\label{hprops}
	h_{i}(t)=\begin{cases}0 \qquad &t\leq0\\
	1 & t\geq1.
 \end{cases}
\end{align}
The interaction Hamiltonian that governs the interaction between the detector and the quantum field is $H_{\textrm{int}}=\alpha\,\hat{\mu}(\tau)\chi(\tau)\,\hat{\phi}(x(\tau))$, where $\alpha$ is a small coupling constant, $\hat{\mu}(\tau)$ is the detector's monopole moment operator and $\chi(\tau)$ is the switching function. Hence, for a switching function of the form (\ref{eq:switchdef}), it is clear that whenever $h_{1}$ or $h_{2}$ are zero, there is no interaction. The interaction is switched on as $h_{1}$ goes from zero to one, which happens during the interval $(\tau_{0}-\delta,\tau_{0})$. Similarly, the interaction is switched off as $h_{2}$ takes on values from one to zero, which happens during the interval $(\tau,\tau+\delta)$. The interaction time then is $\Delta\tau=\tau-\tau_{0}$. The constant $\delta$ measures how rapidly the interaction is switched on and off. A profile of a particular switching function is shown in Fig.~\ref{fig:onoff}.

If we now assume that the interaction time $\Delta\tau$ is much larger than the time interval over which the interaction is switched on or off, then $\delta/\Delta\tau$ is a natural small parameter in which we can expand. The response function expanded in this parameter gives \cite{Satz2007}
\begin{align}
	\label{pert0}
	{\cal F}(E)&=2\int_{\tau_{0}}^{\tau}du\int_{0}^{u-\tau_{0}}ds\left(\cos (E s)W(u,u-s)+\frac{1}{4\pi^{2}s^{2}}\right)
	\nn\\&+\frac{1}{2\pi^{2}}\ln\left(\frac{\Delta\tau}{\delta}\right)-\frac{E}{4\pi}\Delta\tau+C_{1}+{\cal O}\left(\frac{\delta}{\Delta\tau}\right)
	,\end{align}
for some constant $C_1$. In the sharp-switching limit $\delta\to0$, we see an explicit logarithmic divergence in the response function.\footnote{The limit $\delta\to0$ is taken assuming that the detection time $\Delta\tau$ is finite. It is worth noting, however, that the perturbation \eqref{pert0} is also valid for finite $\delta$ and large detection time $\Delta\tau$.} However, differentiating with respect to proper time $\tau$ gives the expansion for the \emph{transition rate}
\begin{align}
	{\cal \dot{F}}_{\tau}(E)&=2\int_{0}^{\Delta\tau}ds\left(\cos(E s)W(\tau,\tau-s)+\frac{1}{4\pi^{2}s^{2}}\right)
	\nn\\&+\frac{1}{2\pi^{2}\Delta\tau}-\frac{E}{4\pi}+{\cal O}\left(\frac{\delta}{(\Delta\tau)^{2}}\right).
\end{align}
Finally, the sharp-switching limit of this expression is evidently regular and gives Eq.~(\ref{eq:transitionratesharp}). 
  
			\section{Large energy gap limit}
			\label{sec:Limit}
  
		Let us rewrite the transition rate as follows
		\begin{align}
			\dot{ \mathcal{F}}_{\tau}(E)=
			\frac{1}{2\pi^{2}}\int_{0}^{\Delta\tau}ds\,\left(\cos (E s)\left[-\frac{1}{a(\tau)a(\tau-s)\left[\eta(\tau)-\eta(\tau-s)\right]^{2}}+\frac{1}{s^{2}}\right]\right)\nonumber\\
			-\frac{1}{2\pi^{2}}\int_{0}^{\Delta\tau}ds\,\left(\frac{\cos(E s)-1}{s^{2}}\right)
			-\frac{E}{4\pi}+\frac{1}{2\pi^{2}\Delta\tau}.
		\end{align}
		The second integral can be computed and expanded in a large $E>0$ series yielding
		\begin{align}
			\dot{ \mathcal{F}}_{\tau}(E)=
			\frac{1}{2\pi^{2}}\int_{0}^{\Delta\tau}ds\,\left(\cos(E s)\left[-\frac{1}{a(\tau)a(\tau-s)\left[\eta(\tau)-\eta(\tau-s)\right]^{2}}+\frac{1}{s^{2}}\right]\right)\nonumber\\
			-\frac{1}{2\pi^{2}}\frac{\sin(E\,\Delta\tau)}{E\,\Delta\tau^{2}}+\mathcal{O}(E^{-2}).
		\end{align}
		If we now integrate the first term by parts, we obtain
		\begin{align}
			\dot{ \mathcal{F}}_{\tau}(E)&=
			-\frac{1}{2\pi^{2}}\frac{\sin(E\Delta\tau)}{E}\frac{1}{a(\tau)a(\tau_{0})[\eta(\tau)-\eta(\tau_{0})]^{2}}\nonumber\\
			& +\frac{1}{2\pi^{2}E}\int_{0}^{\Delta\tau}ds\,\sin(E s)\frac{\partial}{\partial s}\left[-\frac{1}{a(\tau)a(\tau-s)\left[\eta(\tau)-\eta(\tau-s)\right]^{2}}+\frac{1}{s^{2}}\right]\nonumber\\
			&+\mathcal{O}(E^{-2}).
		\end{align}
		By integrating by parts again, it is clear that the term containing the integral is $\mathcal{O}(E^{-2})$, and so to leading order in the large energy gap approximation, we have
		\begin{align}
			\dot{ \mathcal{F}}_{\tau}(E)&=
			-\frac{1}{2\pi^{2}}\frac{\sin(E\Delta\tau)}{E}\frac{1}{a(\tau)a(\tau_{0})[\eta(\tau)-\eta(\tau_{0})]^{2}}+\mathcal{O}(E^{-2}).
		\end{align}
	By multiplying across by $(t_{m}-t_{r})$, the corresponding approximation in terms of the dimensionless quantities is simply
 \begin{align}
			\dot{ \tilde{\mathcal{F}}}_{\tau}(E)&=
			-\frac{1}{2\pi^{2}}\frac{\sin(\tilde{E}\tilde{\Delta\tau})}{\tilde{E}}\frac{1}{a(\tilde{\tau})a(\tilde{\tau}_{0})[\tilde{\eta}(\tilde{\tau})-\tilde{\eta}(\tilde{\tau}_{0})]^{2}}+\mathcal{O}(\tilde{E}^{-2}).
		\end{align}
		A very similar calculation for the limit of large negative energy gap yields
		\begin{align}
			\dot{ \tilde{\mathcal{F}}}_{\tau}(E)&=
			\frac{|\tilde{E}|}{2\pi}-\frac{1}{2\pi^{2}}\frac{\sin(|\tilde{E}|\Delta\tilde{\tau})}{|\tilde{E}|}\frac{1}{a(\tilde{\tau})a(\tilde{\tau}_{0})[\tilde{\eta}(\tilde{\tau})-\tilde{\eta}(\tilde{\tau}_{0})]^{2}}+\mathcal{O}(|\tilde{E}|^{-2}).
		\end{align}

			\bibliographystyle{unsrt}
			\bibliography{allcitations}
		\end{document}